\documentclass[11pt, A4paper]{article}
\usepackage[utf8]{inputenc}
\usepackage[margin=1in]{geometry}
\usepackage[dvipsnames]{xcolor}
\usepackage{times}
\usepackage{amsmath}
\usepackage{amsfonts}
\usepackage{authblk}
\usepackage{graphicx, subfigure}
\usepackage{indentfirst}
\usepackage{amsthm}
\usepackage{lscape}
\usepackage[title,toc,titletoc]{appendix}
\DeclareMathOperator*{\argmax}{arg\,max}
\DeclareMathOperator*{\argmin}{arg\,min}
\usepackage{hyperref}
\hypersetup{colorlinks, urlcolor=black, linkcolor=blue, citecolor=blue}
\usepackage{amssymb, bbm, bm}
\usepackage{setspace}
\usepackage{enumerate}
\usepackage[ruled,vlined]{algorithm2e}
\usepackage{natbib}
\usepackage{booktabs, array}
\usepackage{siunitx} 
\usepackage{color, soul, colortbl}
\usepackage[font=footnotesize,labelfont=bf]{caption}
\usepackage[section]{placeins}
\definecolor{LightCyan}{rgb}{0.88,1,1}

\SetKwRepeat{Do}{do}{while}
\theoremstyle{plain}
\newtheorem{theorem}{Theorem}
\numberwithin{theorem}{section}
\numberwithin{equation}{section}

\newtheorem{lemma}[theorem]{Lemma}

\newcommand{\comment}[1]{}
\allowdisplaybreaks
\usepackage{array}
\newcolumntype{H}{>{\setbox0=\hbox\bgroup}c<{\egroup}@{}}
\title{Robust Maximum L$q$-Likelihood Covariance Estimation \\for Replicated Spatial Data}
\author{Sihan Chen, Joydeep Chowdhury, and Marc G. Genton \\
\small{\textit{Statistics Program, King Abdullah University of Science and Technology}} \\
\small{\{sihan.chen, joydeep.chowdhury, marc.genton\}@kaust.edu.sa}}
\date{}

\begin{document}
\maketitle

\begin{abstract}
  Parameter estimation with the maximum $L_q$-likelihood estimator (ML$q$E) is an alternative to the maximum likelihood estimator (MLE) that considers the $q$-th power of the likelihood values for some $0<q<1$. In this method, extreme values are down-weighted because of their lower likelihood values, which yields robust estimates. In this work, we study the properties of the ML$q$E for spatial data with replicates. We investigate the asymptotic properties of the ML$q$E for Gaussian random fields with a Matérn covariance function, and carry out simulation studies to investigate the numerical performance of the ML$q$E. We show that it can provide more robust and stable estimation results when some of the replicates in the spatial data contain outliers. In addition, we develop a mechanism to find the optimal choice of the hyper-parameter $q$ for the ML$q$E. The robustness of our approach is further verified on a United States precipitation dataset. Compared with other robust methods for spatial data, our proposal is more intuitive and easier to understand, yet it performs well when dealing with datasets containing outliers.
\end{abstract}

\section{Introduction}
\label{sec:intro}

The Maximum Likelihood Estimator (MLE) has been one of the most powerful statistical methods widely used in various domains (\citealp{aldrich1997ra}). Its applications have been extended across various fields, including econometrics (\citealp{greene1980maximum,cramer1989econometric}), genetics (\citealp{shaw1987maximum,beerli2006comparison}) and geography (\citealp{elhorst2005unconditional,ree2008maximum}). The MLE has gained popularity for its asymptotic properties, such as consistency and efficiency (see, e.g., \citealp{casella2021statistical}). It also plays an important role in the analysis of spatial and spatio-temporal data (\citealp{mardia1984maximum,moller2008handbook}), and in recent years, some tools have been developed to efficiently compute the MLE for large-scale spatial data, for example \texttt{ExaGeoStat} (\citealp{exageostat}) and \texttt{GpGp} (\citealp{katzfuss2021general}). 

However, when using the MLE with datasets containing outliers, the likelihood evaluation might be largely distorted by these extreme values, and this can lead to inaccurate estimation results (\citealp{chen2014comparison}). The maximum $L_q$-Likelihood estimator (ML$q$E) proposed by \cite{lq} provides an effective way to address this problem. Unlike the MLE which calculates the direct summation of the log-likelihood values at each data point, the ML$q$E sums the $q$-th power of the likelihood values for  some $q<1$, which down-weights the extreme values and leads to robust estimation. \cite{lq} and \cite{ferrari} established the asymptotic normality of the ML$q$E under some regularity conditions and provided the expression of its asymptotic variance. 

In this work, our aim is to employ the ML$q$E method to estimate the covariance parameters for replicated spatial data. We consider the replications of spatial data as independent realisations of a random field over the spatial domain, and estimate the parameters of the covariance function of the random field using the ML$q$E. The outliers in this work are defined in the following way. Outliers are considered to be replicates of the random field whose dependence structures are largely different from the rest of the sample. The fraction of outliers in the data depends highly on the application and can vary significantly; therefore, in the simulation experiments of this work, we consider several different settings, where the fraction of contaminated data in the datasets varies from 1\% to 20\%.
We focus on Gaussian random fields with the Matérn covariance function, which is one of the most influential models for spatial statistics. The Matérn covariance function was proposed by \cite{matern1960spatial}, and became popular later after the work of \cite{handcock1993bayesian}. It is used in many well-known tools in \texttt{R}, such as \texttt{RandomFields} (\citealp{schlather2013randomfields}) and \texttt{fields} (\citealp{nychka2021fields}). A comprehensive review of the application of the Matérn covariance function in spatial statistics and many other related research fields can be found in \cite{porcu2023mat}. There are several different widely accepted ways to parameterise the Matérn covariance function; see \cite{wang2023parameterization} for a comprehensive comparison of the three most popular parameterisations. In this paper, we adopt the following formulation of the Matérn covariance function, which is the first type of parametrisation in \cite{wang2023parameterization}:
\begin{align}
    \label{matern}
    \mathcal{M}( h; \boldsymbol{\theta} ) = \frac{\sigma^2}{\Gamma( \nu ) 2^{\nu - 1}} \left( \frac{h}{\beta} \right)^{\nu}\mathcal{K}_\nu\left( \frac{h}{\beta} \right) .
\end{align}
Here, $h$ is the distance between a pair of spatial locations, and $\boldsymbol{\theta} = ( \sigma^2, \beta, \nu )^\top$ is the parameter vector that we aim to estimate in a robust way with our proposed method. Its components $\sigma^2, \beta, \nu$ are the variance, range and smoothness parameters,  respectively, and $\mathcal{K}_\nu( \cdot )$ is the modified Bessel function of the second kind of order $\nu$. 

In the literature, there have been various attempts to apply robust statistical methods on spatial data. For example, \cite{cressie1980robust} proposed a robust procedure for estimating the variogram of spatial data, and \cite{hawkins1984robust} also proposed a robust kriging method based on their previous work. \cite{genton1998highly} proposed a highly robust variogram estimator and compared it with the one in \cite{cressie1980robust}, while showing that the latter is actually not strictly robust but only less non-robust than the empirical variogram estimator. A comprehensive review of classical robust methods for spatial data can be found in \cite{lark2000comparison}. \cite{marchant2007robust} used the residual maximum likelihood for robust estimation of the variogram, which was based on the robustified MLE proposed in \cite{richardson1995robust} for linear mixed models. The latest approach proposed by \cite{kunsch2013robust} is a robust method to estimate the external drift and the variogram of spatial data. Their main idea is to replace the residual function with another bounded function of the observations, so that the influence of outliers in the data can be bounded. Unfortunately, their method is not very practical because it is not able to deal with large-scale datasets. Indeed, in their numerical studies they only used datasets limited to a few hundreds of locations, which is often not sufficient because the scale of modern spatial datasets is getting very large. Hence, a robust likelihood-based method for estimating the parameters of the spatial covariance function for large-scale datasets is lacking.





\begin{figure}[h!]
    \centering
    \includegraphics[width=0.8\linewidth]{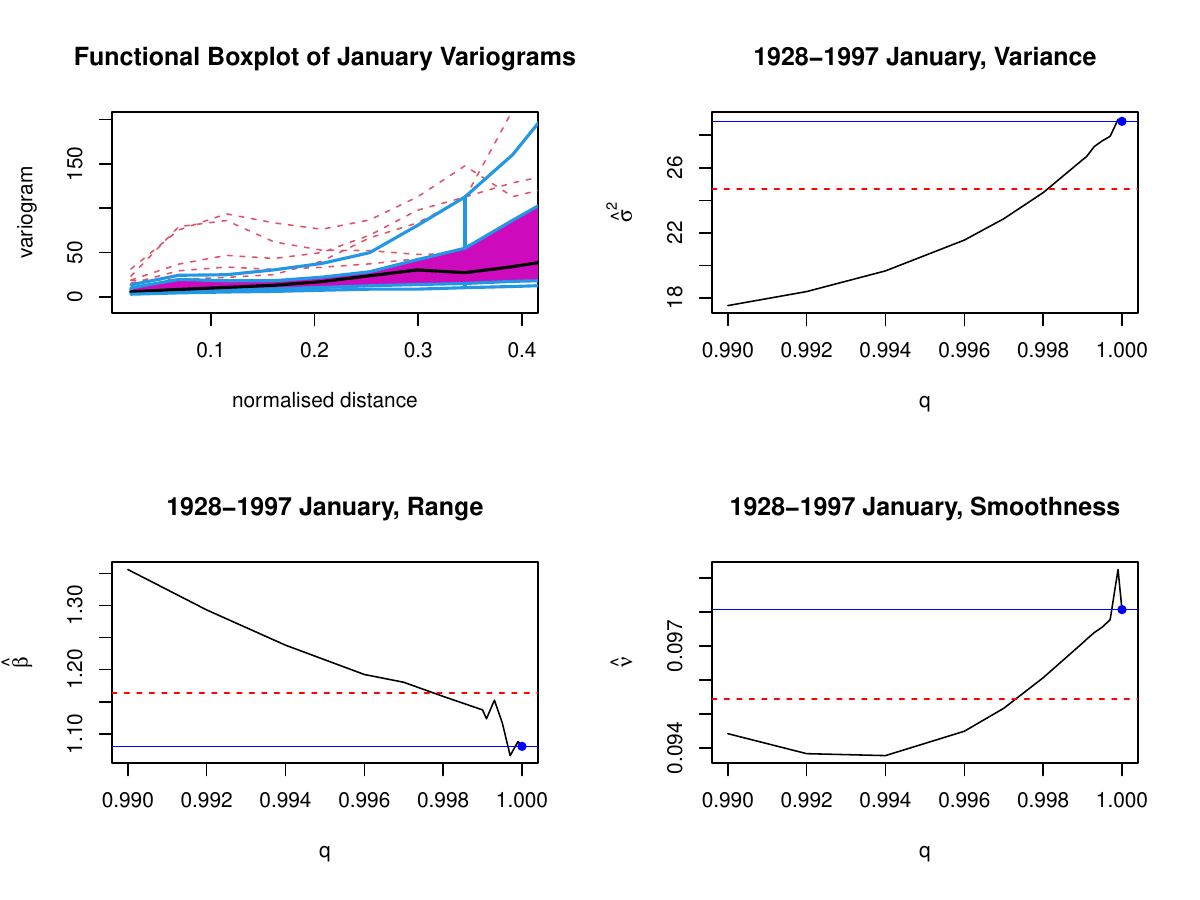}
    \caption{\footnotesize Plots from the motivating example of US precipitation data. In the first sub-figure we present the functional boxplot of the variograms of the January precipitation data, which we use to identify the outliers in the dataset. The remaining three sub-figures present MLE and ML$q$E estimation results of the three parameters, variance (${\sigma}^2$), range (${\beta}$) and smoothness (${\nu}$), with the January data from 1928 to 1997. The black curves represent the ML$q$E results with different values of $q$, and when $q=1$ (indicated by the blue dot on the curve), the ML$q$E is the same as the MLE. The blue horizontal lines and the red dashed lines denote the MLE results with and without the outliers in the dataset. 
    }
    \label{real4}
\end{figure}

We first demonstrate how the ML$q$E diminishes the influence of outliers in the dataset on parameter estimation results using a motivating example of precipitation data from the United States retrieved from \url{https://www.image.ucar.edu/Data/US.monthly.met/USmonthlyMet.shtml}. The detailed definition of the ML$q$E can be found in Section \ref{sec:method}. The dataset that we consider here contains hourly precipitation data from 1895 to 1997 recorded at thousands of observation stations throughout the country. In this example, we consider the average hourly precipitation in the month of January from 1928 to 1997 recorded at 621 different monitoring stations located in the contiguous United States. The coordinates of these locations are normalised to be within the $2$-dimensional unit square. We first conduct the parameter estimation using the MLE and the ML$q$E with several different values of the hyper-parameter $q$, and then try to remove the years in the data with the most outliers and redo the parameter estimation using the MLE, to see whether the MLE after removing the outliers is closer to the ML$q$E without removing the outliers for some $q<1$. For the average January data for each year, we first estimate the empirical variogram of the data, which is an empirical function that describes the spatial dependence of the random field, using the \texttt{R} package \texttt{geoR} (\citealp{ribeiro2007geor}); next, we find out the years for which the January data behave differently from other years via a functional boxplot of the corresponding empirical variogram, using the \texttt{R} command \texttt{fbplot} (\citealp{sun2011functional}). The resulting functional boxplot for the January data of all the 70 years is shown in the first sub-figure of Figure \ref{real4}, where the dashed-curves represent the outliers identified among the 70 empirical variograms. Here we identify 5 outliers in total. Additionally, in the same plot, the black curve is the median of the variograms, and the purple area represents the 50\% central region. In the last three sub-figures of Figure \ref{real4}, the blue horizontal lines represent the MLE results with the original data, and the red dashed lines are the MLE results after removing the 5 outliers from the dataset, while the black curves are the ML$q$E results without removing the outliers. The blue dots in those sub-figures are the ML$q$E values for $q=1$, which coincide with the MLE values. We can see that the red dashed line intersects the black curve at a $q<1$, which means that the existence of outliers significantly affects the parameter estimation results, and the ML$q$E is able to reduce their effect with a $q<1$. It suggests that applying the ML$q$E for some $q<1$ is able to provide parameter estimation results that automatically diminish the effect of outliers, without the need to identify and remove outliers manually in the dataset beforehand. 

In this paper, we first investigate the asymptotic properties of the ML$q$E in the scenario of Gaussian random fields with Matérn covariance function. Then, we carry out simulation studies to investigate the numerical performance of the ML$q$E with synthetic spatial data, which demonstrate that the ML$q$E is able to provide a more robust and stable parameter estimation in the presence of outliers. This observation is also corroborated by experiments with real data. Next, we develop a methodology to choose the optimal value of the hyper-parameter $q$ for the ML$q$E. Since spatial data are often large-scale in practice, we implement our method using the software \texttt{ExaGeoStat}, which provides a high-performance computing framework for spatial and spatio-temporal data, making our implementation capable of dealing with large-scale data efficiently. 

The remainder of this paper is organised as follows. Section \ref{sec:method} describes the proposed methodology and provides some of its theoretical properties on Gaussian random fields, with some of the corresponding proofs placed in Appendix~\ref{sec:proof}. We also present our procedure of tuning the hyper-parameter $q$ for the ML$q$E, as well as our algorithm for computing the L$q$-likelihood. In Section \ref{sec:simulation}, we present the data generation method and numerical results from simulation studies for synthetic data, and compare the behaviour of the ML$q$E and the MLE on both clean data and contaminated data, while some additional experimental results can be found in the supplementary materials. The experiments on our mechanism for choosing the optimal value of $q$ are included in Section \ref{sec:simulation} as well. Section \ref{sec:realdata} expands the application of the ML$q$E on the precipitation data in the United States. Conclusions and discussions are presented in Section \ref{sec:conclusion}. 

\section{Method}
\label{sec:method}
\subsection{Maximum $L_q$-Likelihood Estimator}
Let $f_1$ and $f_2$ be two probability density functions and $X$ be a random variable whose density exists. The $q$-entropy (\citealp{tsallis1988possible}) of $f_2$ with respect to $f_1$ is defined as
\begin{equation}
    \mathcal{H}_q(f_1, f_2)=-\mathbb{E}_{f_1}\{L_q[f_2(X)]\},\quad q>0,
\label{qentropy}
\end{equation}
where 
\begin{equation}
\label{lq}
    L_q(u)=
    \begin{cases}
    \log u,\quad\text{if}\quad q=1; \\
    \left(u^{1-q}-1\right)/(1-q),\quad\text{otherwise}.
    \end{cases}
\end{equation}

The Maximum $L_q$-Likelihood Estimator (ML$q$E) is defined through the empirical $q$-entropy, where we replace $f_1$ and $f_2$ in (\ref{qentropy}) by the empirical density and the density of the parametric model, respectively. Consider an i.i.d. sample $X_1, \dots, X_n$ from the distribution $g(x;\boldsymbol{\theta}_0)$ with parameter $\boldsymbol{\theta}_0\in\boldsymbol{\Theta} \subseteq \mathbb{R}^p$ for some integer $p\geq1$. The ML$q$E for the true parameter $\boldsymbol{\theta}_0$ is formulated as 
\begin{equation}
\hat{\boldsymbol{\theta}}=\argmax_{\boldsymbol{\theta}\in\boldsymbol{\Theta}}\sum_{i=1}^nL_q\{g(X_i; \boldsymbol{\theta})\},\quad 0<q\leq1.
\label{mlqe}
\end{equation}

From this definition we can see that, when $q=1$, the ML$q$E is exactly the same as the MLE, because in this case the function $L_q$ is nothing but the logarithmic function. In addition, it can be shown that, when $q\to1$, if the ML$q$E $\hat{\boldsymbol{\theta}}$ exists, it approaches the MLE.



\subsection{ML$q$E for Gaussian Random Fields}
\label{sec:mlqegf}

    

The main objective of this work is to apply the ML$q$E to spatial data with replicates. We consider $m$ sets of independent realisations of an identical zero-mean Gaussian random field, with $m>1$, denoted by $\{Z_i(\mathbf{s)}: \mathbf{s}\in\mathbb{R}^d\}$, where $i=1, \ldots, m$ and $d\in\mathbb{Z}^+$. Here, $d$ denotes the dimension of the locations. For most geographical and environmental datasets, we have $d=2$, and this will be the setting for all the datasets that we use in the numerical experiments of this work as well. The spatial covariance function of each of the $Z_i$s is assumed to be the Matérn covariance function shown in (\ref{matern}). We consider the case in which all the $m$ realisations of the random fields, denoted by $\{\mathbf{Z}_i, i=1, \dots, m\}$, are recorded on a fixed set of $n$ locations, where $n>1$. Therefore, each $\mathbf{Z}_i$ is an $n$-dimensional vector with identical zero-mean multivariate Gaussian distribution with common probability density
\begin{align}
\label{density}
f( \mathbf{z} ; \boldsymbol{\theta} ) = \frac{1}{\left|2 \pi\mathbf{\Sigma}_{\mathcal{M}}\right| ^{1/2}} \exp\left(- \frac{1}{2} \mathbf{z}^\top \mathbf{\Sigma}_{\mathcal{M}}^{-1} \mathbf{z}\right) ,\quad  \mathbf{z}\in\mathbb{R}^n,
\end{align}
where $\mathbf{\Sigma}_{\mathcal{M}}$ is the $n \times n$ covariance matrix obtained from the Matérn covariance function $\mathcal{M}( h; \boldsymbol{\theta} )$ in (\ref{matern}) based on the $n$ locations. 

It follows that the ML$q$E $\hat{\boldsymbol{\theta}}$ defined in (\ref{mlqe}) for $\{\mathbf{Z}_i, i=1, \dots, m\}$ can be expressed as
\begin{equation*}
\hat{\boldsymbol{\theta}}=\argmax_{\boldsymbol{\theta}\in\boldsymbol{\Theta}}\sum_{i=1}^mL_q\{f(\mathbf{Z}_i; \boldsymbol{\theta})\},\quad 0<q\leq1,
\end{equation*}
where the function $f$ is as in (\ref{density}) and $L_q$ is as in (\ref{lq}).

Define
\begin{align*}
\mathbf{U}\left( \mathbf{z}; \boldsymbol{\theta} \right)
= \frac{\partial}{\partial \boldsymbol{\theta}} \log f\left( \mathbf{z}; \boldsymbol{\theta} \right).
\end{align*}
Then for $\{\mathbf{Z}_i, i=1, \dots, m\}$, the ML$q$E $\hat{\boldsymbol{\theta}}$ can alternatively be represented by the solution of the following equation:
\begin{equation}
\label{mlqequation}
    \sum_{i=1}^m\mathbf{U}( \mathbf{Z}_i; \hat{\boldsymbol{\theta}})f( \mathbf{Z}_i; \hat{\boldsymbol{\theta}} )^{1-q}=\mathbf{0},\quad 0<q\leq1.
\end{equation}


\subsection{Asymptotic Properties}
\label{sec:asymptotic}

In this subsection, we prove the consistency and derive the asymptotic variance of the ML$q$E for Gaussian random fields with Matérn covariance. 

Note that the density function (\ref{density}) can be rewritten as
\begin{align*}
f( \mathbf{z} ; \boldsymbol{\theta} ) = \exp\left[ \left\{\boldsymbol{\eta}\left(\boldsymbol{\theta}\right)\right\}^\top \mathbf{b}\left(\mathbf{z}\right) - A\left(\boldsymbol{\theta}\right) \right] ,
\end{align*}
where we define
\begin{align*}
\mathbf{b}\left(\mathbf{z}\right)
=
\begin{bmatrix}
\mathbf{z} \\
\text{vec}\left( {\mathbf{z} \mathbf{z}^\top} \right) 
\end{bmatrix},\qquad
 \boldsymbol{\eta}\left(\boldsymbol{\theta}\right)
= \begin{bmatrix}
\mathbf{0}_{n} \\
\text{vec}\left( - \frac{1}{2} \mathbf{\Sigma}_{\mathcal{M}}^{-1} \right) 
\end{bmatrix},
\end{align*}
and $A\left(\boldsymbol{\theta}\right)$ is a normalising constant. Recall that according to (\ref{matern}), the dimension of $\boldsymbol{\theta}$ is 3, which is smaller than the dimension of the vector $\boldsymbol{\eta}\left(\boldsymbol{\theta}\right)$ for any integer $n>1$. Therefore, $f( \mathbf{z} ; \boldsymbol{\theta} )$ belongs to a curved exponential family, see p.~25 and Note 10.6 on p.~79 in \cite{lehmann2006theory}.

In Section 3 of \cite{lq}, the asymptotic distribution theory for ML$q$E was derived for full-rank exponential families. However, the parameters for curved exponential families estimated via the ML$q$E also follow similar asymptotic normality as derived in Theorem 3.2 in \cite{lq}, and further generalised in Theorem 4.2 in the same article. Using these results, we have the following lemma and theorem.

\begin{lemma}
Let $q_m >0$ satisfy $q_m \to 1$ as $m \to \infty$, and let the value of the underlying parameter $\boldsymbol{\theta}_0$ of the sample be an interior point of the parameter space $\boldsymbol{\Theta}$, which is compact. Then, the probability that Equation \eqref{mlqequation} has a unique solution $\hat{\boldsymbol{\theta}}$ converges to 1 as $m \to \infty$ and $\hat{\boldsymbol{\theta}} \stackrel{P}{\longrightarrow} \boldsymbol{\theta}_0$ as $m \to \infty$.
\end{lemma}
\begin{proof}
This lemma follows directly from Theorem 3.2 in \cite{lq}. 
\end{proof}

Next, the following theorem states asymptotic Gaussianity of the ML$q$E along with the form of the asymptotic dispersion matrix. Before stating the theorem, we first define some necessary quantities for it.
Let 
\begin{align}
\label{uv}
\mathbf{U}^*_n( \mathbf{Z}; \boldsymbol{\theta}, q ) =& \left(\frac{1}{(2\pi)^{\frac{n}{2}}|\mathbf{\Sigma}_{\mathcal{M}}|^{\frac{1}{2}}}\right)^{1-q}\exp\left\{-\frac{1-q}{2}\left(\mathbf{Z}^\top\mathbf{\Sigma}_{\mathcal{M}}^{-1}\mathbf{Z}\right)\right\} \nonumber\\
&\times\left[\frac{1}{2}\mathbf{Z}^\top\mathbf{\Sigma}_{\mathcal{M}}^{-1}\frac{\partial\mathbf{\Sigma}_{\mathcal{M}}}{\partial\boldsymbol{\theta}}\mathbf{\Sigma}_{\mathcal{M}}^{-1}\mathbf{Z} - \frac{1}{2}\text{tr}\left(\mathbf{\Sigma}_{\mathcal{M}}^{-1}\frac{\partial}{\partial\boldsymbol{\theta}}\mathbf{\Sigma}_{\mathcal{M}}\right)\right] ,\\
 \mathbf{V}^*_n( \mathbf{Z}; \boldsymbol{\theta}, q ) 
=& (1-q)\left(\frac{1}{(2\pi)^{\frac{n}{2}}|\mathbf{\Sigma}_{\mathcal{M}}|^{\frac{1}{2}}}\right)^{1-q}\exp\left(-\frac{1-q}{2}\left(\mathbf{Z}^\top\mathbf{\Sigma}_{\mathcal{M}}^{-1}\mathbf{Z}\right)\right)\nonumber\\ &\times 
    \left[\frac{1}{2}\mathbf{Z}^\top\mathbf{\Sigma}_{\mathcal{M}}^{-1}\frac{\partial\mathbf{\Sigma}_{\mathcal{M}}}{\partial\boldsymbol{\theta}}\mathbf{\Sigma}_{\mathcal{M}}^{-1}\mathbf{Z}-\frac{1}{2}\text{tr}\left(\mathbf{\Sigma}_{\mathcal{M}}^{-1}\frac{\partial}{\partial\boldsymbol{\theta}}\mathbf{\Sigma}_{\mathcal{M}}\right)\right]^2 \nonumber\\
    &+ \left(\frac{1}{(2\pi)^{\frac{n}{2}}|\mathbf{\Sigma}_{\mathcal{M}}|^{\frac{1}{2}}}\right)^{1-q}\exp\left(-\frac{1-q}{2}\left(\mathbf{Z}^\top\mathbf{\Sigma}_{\mathcal{M}}^{-1}\mathbf{Z}\right)\right)\nonumber\\ &\times 
    \left\{\left[\frac{1}{2}\mathbf{Z}^\top\mathbf{\Sigma}_{\mathcal{M}}^{-1}\left[\frac{\partial^2\mathbf{\Sigma}_{\mathcal{M}}}{\partial\boldsymbol{\theta}^2}-2\frac{\partial\mathbf{\Sigma}_{\mathcal{M}}}{\partial\boldsymbol{\theta}}\mathbf{\Sigma}_{\mathcal{M}}^{-1}\frac{\partial\mathbf{\Sigma}_{\mathcal{M}}}{\partial\boldsymbol{\theta}}\right]\mathbf{\Sigma}_{\mathcal{M}}^{-1}\mathbf{Z}\right]\right.\nonumber\\ &\left.-\frac{1}{2}\text{tr}\left(-\mathbf{\Sigma}_{\mathcal{M}}^{-1}\frac{\partial\mathbf{\Sigma}_{\mathcal{M}}}{\partial\boldsymbol{\theta}}\mathbf{\Sigma}_{\mathcal{M}}^{-1}\frac{\partial\mathbf{\Sigma}_{\mathcal{M}}}{\partial\boldsymbol{\theta}}+\mathbf{\Sigma}_{\mathcal{M}}^{-1}\frac{\partial^2\mathbf{\Sigma}_{\mathcal{M}}}{\partial\boldsymbol{\theta}^2}\right)\right\},
\end{align}
where $\mathbf{\Sigma}_{\mathcal{M}}$, $\frac{\partial}{\partial\boldsymbol{\theta}}\mathbf{\Sigma}_{\mathcal{M}}$ and $\frac{\partial^2}{\partial\boldsymbol{\theta}^2}\mathbf{\Sigma}_{\mathcal{M}}$ are functions of $\boldsymbol{\theta}$, but we drop the argument $\boldsymbol{\theta}$ for brevity. Here for any vector $\boldsymbol{a}$, we denote the operation $\boldsymbol{a}^{\otimes2}=\boldsymbol{a}\boldsymbol{a}^\top$; in addition, $\frac{\partial}{\partial\boldsymbol{\theta}}\mathbf{\Sigma}_{\mathcal{M}}$ is a $p$-dimensional vector of matrices, and $\frac{\partial^2}{\partial\boldsymbol{\theta}^2}\mathbf{\Sigma}_{\mathcal{M}}$ is a $p\times p$ matrix of matrices, where $p$ is the dimension of $\boldsymbol{\theta}$. Every element of $\frac{\partial}{\partial\boldsymbol{\theta}}\mathbf{\Sigma}_{\mathcal{M}}$ and $\frac{\partial^2}{\partial\boldsymbol{\theta}^2}\mathbf{\Sigma}_{\mathcal{M}}$ is an $n\times n$ matrix, and $\frac{\partial}{\partial\boldsymbol{\theta}}\mathbf{\Sigma}_{\mathcal{M}}$ and $\frac{\partial^2}{\partial\boldsymbol{\theta}^2}\mathbf{\Sigma}_{\mathcal{M}}$ are defined as follows:
\begin{align*}
    \frac{\partial}{\partial\boldsymbol{\theta}}\mathbf{\Sigma}_{\mathcal{M}}=\left(\frac{\partial}{\partial\theta_1}\mathbf{\Sigma}_{\mathcal{M}}, \dots, \frac{\partial}{\partial\theta_p}\mathbf{\Sigma}_{\mathcal{M}}\right)^\top; \,\,
    \frac{\partial^2}{\partial\boldsymbol{\theta}^2}\mathbf{\Sigma}_{\mathcal{M}}=
    \begin{bmatrix}
        \frac{\partial^2}{\partial\theta_1^2}\mathbf{\Sigma}_{\mathcal{M}} & \frac{\partial^2}{\partial\theta_1\theta_2}\mathbf{\Sigma}_{\mathcal{M}} & \dots & \frac{\partial^2}{\partial\theta_1\theta_p}\mathbf{\Sigma}_{\mathcal{M}} \\
        \vdots & \vdots & \ddots & \vdots \\
        \frac{\partial^2}{\partial\theta_p\theta_1}\mathbf{\Sigma}_{\mathcal{M}} & \frac{\partial^2}{\partial\theta_p\theta_2}\mathbf{\Sigma}_{\mathcal{M}} & \dots & \frac{\partial^2}{\partial\theta_p^2}\mathbf{\Sigma}_{\mathcal{M}}
    \end{bmatrix}.
\end{align*}
The matrix operations involving $\frac{\partial}{\partial\boldsymbol{\theta}}\mathbf{\Sigma}_{\mathcal{M}}$ and $\frac{\partial^2}{\partial\boldsymbol{\theta}^2}\mathbf{\Sigma}_{\mathcal{M}}$ are carried out element-wise over the individual $n \times n$ matrix elements, so that $\mathbf{U}^*_n( \mathbf{Z}; \boldsymbol{\theta}, q )$ is a $p$-dimensional vector and $\mathbf{V}^*_n( \mathbf{Z}; \boldsymbol{\theta}, q )$ is a $p\times p$ matrix. Define $\boldsymbol{\theta}_m^*$ such that it satisfies
$\text{E}_{\boldsymbol{\theta}_0}\left[ \mathbf{U}^*_n( \mathbf{Z}; \boldsymbol{\theta}_m^*, q_m ) \right] = \mathbf{0}$.
Also define $\mathbf{J}_m=\text{E}_{\boldsymbol{\theta}_0}\left[ \mathbf{V}^*_n( \mathbf{Z}; \boldsymbol{\theta}_m^*, q_m ) \right]$, $\mathbf{K}_m= \text{E}_{\boldsymbol{\theta}_0}\left[ \mathbf{U}^*_n( \mathbf{Z}; \boldsymbol{\theta}_m^*, q_m ) \mathbf{U}^*_n( \mathbf{Z}; \boldsymbol{\theta}_m^*, q_m )^\top \right]$.
We have the following theorem:
\begin{theorem}
\label{th1}
Let $q_m \to 1$ as $m \to \infty$, and the parameter value $\boldsymbol{\theta}_0$ of the sample be an interior point of the compact parameter space $\boldsymbol{\Theta}$. Then,
\begin{align*}
\sqrt{m} \left(\mathbf{J}_m^{-1} \mathbf{K}_m \mathbf{J}_m^{-1}\right)^{-1/2} \left( \hat{\boldsymbol{\theta}} - \boldsymbol{\theta}_m^* \right) \stackrel{d}{\longrightarrow} N_p( \mathbf{0}_p, \mathbf{I}_p ) \text{ as } m \to \infty .
\end{align*}
If $\sqrt{m} (q_m - 1 ) \to 0$ as $m \to \infty$, then we have
\begin{align*}
\sqrt{m} \left(\mathbf{J}_m^{-1} \mathbf{K}_m \mathbf{J}_m^{-1}\right)^{-1/2} \left( \hat{\boldsymbol{\theta}} - \boldsymbol{\theta}_0 \right) \stackrel{d}{\longrightarrow} N_p( \mathbf{0}_p, \mathbf{I}_p ) \text{ as } m \to \infty .
\end{align*}
\end{theorem}

Here, $\mathbf{0}_p$ and $\mathbf{I}_p$ represent the zero vector of dimension $p$ and the identity matrix of dimension $p\times p$, respectively. The proof of this theorem, as well as the expressions of the derivatives $\frac{\partial}{\partial\boldsymbol{\theta}}\mathbf{\Sigma}_{\mathcal{M}}$ and $\frac{\partial^2}{\partial\boldsymbol{\theta}^2}\mathbf{\Sigma}_{\mathcal{M}}$, are deferred to Appendix~\ref{sec:proof}.


\subsection{Influence Function}
The robustness of the ML$q$E method was explored in \cite{ferrari} for multivariate observations using the influence function. Similar results can be obtained for replicated spatial data in our setup, which we present here. 
\begin{lemma}
Let the influence function for $\hat{\boldsymbol{\theta}}$ be denoted by $\textbf{IF}_q( \mathbf{z}, \boldsymbol{\theta} )$. Then,
\begin{align*}
\textbf{IF}_q( \mathbf{z}, \boldsymbol{\theta} ) = -q^{-1} \mathbf{J}_m^{-1} \mathbf{U}^*_n( \mathbf{z}; \boldsymbol{\theta}, q ) .
\end{align*}
\end{lemma}
The proof of the lemma follows directly from the discussion in Section 3.2 in \cite{ferrari}. Note that
\begin{align*}
\min\{ | \lambda_i | \} \| \mathbf{z} \|^2
\le \left| \mathbf{z}^\top\mathbf{\Sigma}_{\mathcal{M}}^{-1}\frac{\partial\mathbf{\Sigma}_{\mathcal{M}}}{\partial\boldsymbol{\theta}}\mathbf{\Sigma}_{\mathcal{M}}^{-1}\mathbf{z} \right|
\le \max\{ | \lambda_i | \} \| \mathbf{z} \|^2 ,
\end{align*}
where $\{ \lambda_i \}$ are the eigenvalues of $\mathbf{\Sigma}_{\mathcal{M}}^{-1}\frac{\partial\mathbf{\Sigma}_{\mathcal{M}}}{\partial\boldsymbol{\theta}}\mathbf{\Sigma}_{\mathcal{M}}^{-1}$.
Since for any $ q < 1$,
\begin{align*}
\int \| \mathbf{z} \|^2 \exp\left\{-\frac{1-q}{2}\left(\mathbf{z}^\top\mathbf{\Sigma}_{\mathcal{M}}^{-1}\mathbf{z}\right)\right\} \mathrm{d} \mathbf{z} < \infty ,
\end{align*}
as $\exp\left\{-\frac{1-q}{2}\left(\mathbf{z}^\top\mathbf{\Sigma}_{\mathcal{M}}^{-1}\mathbf{z}\right)\right\}$ corresponds to the density of a Gaussian random vector, we have
\begin{align*}
\| \mathbf{z} \|^2 \exp\left\{-\frac{1-q}{2}\left(\mathbf{z}^\top\mathbf{\Sigma}_{\mathcal{M}}^{-1}\mathbf{z}\right)\right\} \to 0
\end{align*}
as $\| \mathbf{z} \| \to \infty$. Therefore, it follows that the influence function $\textbf{IF}_q( \mathbf{z}, \boldsymbol{\theta} )$ is bounded when $q<1$. On the other hand, when $q = 1$, which is the case of the MLE, the exponential term in $\mathbf{U}^*_n( \mathbf{z}; \boldsymbol{\theta}, q )$ vanishes, and 
hence $\textbf{IF}_q( \mathbf{z}, \boldsymbol{\theta} )$ is unbounded. The boundedness can be further verified by the sensitivity curves for the ML$q$E, which are presented in Figures~S1 and S2 in the supplementary materials.

\subsection{Choice of $q$}
\label{sec:choiseofq}
Although smaller values of the hyper-parameter $q$ can reduce the influence of outliers in the data to a larger extent, it also results in higher variance and numerical instability, as we will see from the numerical results in Sections~\ref{sec:nooutlier} and \ref{sec:withoutliers}. Therefore, a mechanism for finding the optimal value of $q$ is desired. The method we apply to tune the hyper-parameter $q$ is inspired by the one introduced in \cite{ribeiro}, with the idea of performing an adequate grid search. The goal is to find a sub-interval of $(0, 1]$ such that all values of $q$ taken from this sub-interval lead to similar estimation results. For this purpose, we need to find a metric that jointly evaluates the three parameters that we estimate. 

The first option is to follow the idea in \cite{ribeiro}, which is to standardise the parameter estimates with their corresponding asymptotic variance, and then sum them up. Define an ordered grid of values for $q$, denoted by $q_0=1>q_1>q_2>\dots>q_K>0$, and for each $q_k$ with $0\leq k\leq K$, denote the corresponding $p$-dimensional ML$q$E by $$\hat{\boldsymbol{\theta}}_{q_k}=\left(\hat{\theta}_{q_k}^1, \dots, \hat{\theta}_{q_k}^p\right)^\top,$$ and define the corresponding vector of standardised estimates $\boldsymbol\zeta_{q_k}$ as $$\boldsymbol{\zeta}_{q_k}=\left(\frac{\hat{\theta}_{q_k}^1}{\sqrt{m}\,\text{se}\left(\hat{\theta}_{q_k}^1\right)}, \dots, \frac{\hat{\theta}_{q_k}^p}{\sqrt{m}\,\text{se}\left(\hat{\theta}_{q_k}^p\right)}\right)^\top,$$ where $m$ is the number of replicates of the spatial data, and $\text{se}(\hat{\theta})$ denotes the asymptotic standard deviation of $\hat{\theta}$ derived in Section \ref{sec:asymptotic}. For $1\leq r\leq p$, $\text{se}(\hat{\theta}_{q_k}^r)$ is the $r$-th diagonal entry of the matrix $$\left(\mathbf{J}_m(\hat{\boldsymbol{\theta}}_{q_k})^{-1} \mathbf{K}_m(\hat{\boldsymbol{\theta}}_{q_k})\mathbf{J}_m(\hat{\boldsymbol{\theta}}_{q_k})^{-1}\right)^{1/2}.$$


Furthermore, we define the standardised quadratic variation (SQV) for each $q_k$ with $1\leq k\leq K$ as $$\text{SQV}_{q_k}=\|\boldsymbol{\zeta}_{q_{k-1}}-\boldsymbol{\zeta}_{q_k}\|/p,$$ and choose the value of $q$ based on this quantity. 

The corresponding algorithm to find the optimal value of $q$, denoted by $q^*$, through SQV, is shown in Algorithm \ref{algq1}. This algorithm basically follows what is proposed in \cite{ribeiro} and aims to find an optimal $q^*$ such that the ML$q$E is stabilised in a neighborhood of $q^*$. However, if no optimal value of $q^*$ can be found in this way, the algorithm still returns $q^*=1$: we may not be able to obtain robustness in this case, so we still choose the MLE to ensure stability. 
\begin{algorithm}[h!]
\footnotesize
\SetAlgoLined
\SetKwInOut{Input}{Input}
\SetKwInOut{Output}{Output}
\smallskip
\Input{An ordered grid for $q$: $q_0=1>q_1>\dots>q_K=q_{min}>0$, a threshold for the SQV: $L>0$, a threshold for the difference between values of $q$: $\epsilon>0$}
\smallskip
\While{$q_0-q_{min}>\epsilon$,}{
\smallskip
Initialise $q^*\gets1$; \\
Calculate $\text{SQV}_{q_k}$ for each $k=1, \dots, K$; \\
\If {$\forall k$, SQV$_{q_k}<L$}{
\smallskip
$q^*\gets q_0$; \\
\textbf{break};
}
Let $k^*$ be the largest integer among $1, \dots, K$ such that $\text{SQV}_{q_{k^*}}\geq L$; \\
Define a new equally spaced grid for $q$: $q_0=q_{k^*}>q_1>\dots>q_K=q_{min}>0$; 
}
\Output{$q^*$}
\caption{Tuning the hyper-parameter $q$ with SQV}
\label{algq1}
\end{algorithm}

However, directly summing up the standardised values of the parameters may not be the best way to select $q$. Another option we consider is to make use of the function $\kappa$ defined as
\begin{equation}
    \kappa(\sigma^2, \beta, \nu)=\sigma^2\beta^{-2\nu},
\label{functionf}
\end{equation}
which was proved in \cite{zhang} to be able to identify Matérn covariance parameters under infill asymptotics. Therefore, compared to the previous option using the SQV, the function $\kappa$ is capable of providing an evaluation of the fitness of the model with the estimated parameters. Moreover, the computation time using the function $\kappa$ is much faster, since computing the SQV requires computing the asymptotic standard errors of the parameters, which is extremely time-consuming, since it involves a lot of complicated large matrix operations.

Similarly to the previous option, for the parameter estimation results using each $q_k$ with $0\leq k\leq K$, we calculate the corresponding value of the function $\kappa$, denoted by $\kappa_{q_k}$. The algorithm for finding the optimal value of $q$, denoted by $q^*$, through the function $\kappa$, is shown in Algorithm \ref{algq2}.
\begin{algorithm}[h!]
\footnotesize
\SetAlgoLined
\SetKwInOut{Input}{Input}
\SetKwInOut{Output}{Output}
\smallskip
\Input{An ordered grid for $q$: $q_0=1>q_1>\dots>q_K=q_{min}>0$, a threshold coefficient: $L>0$, a threshold for the difference between values of $q$: $\epsilon>0$}
\smallskip
\While{$q_0-q_{min}>\epsilon$,}{
\smallskip
Initialise $q^*\gets1$; \\
Calculate $\kappa_{q_k}$ for each $k=0, \dots, K$; \\
Calculate $d\kappa_{q_k}=|\kappa_{q_{k-1}}/\kappa_{q_k}-1|$ for each $k=1, \dots, K$; \\
\If {$\max_k\{d\kappa_{q_k}\}<L \mathcolor{blue}{\cdot} \min_k\{d\kappa_{q_k}\}$}{
\smallskip
$q^*\gets q_0$; \\
\textbf{break};
}
Let $k^*$ be the largest integer among $1, \dots, K$ such that $d\kappa_{q_{k^*}}\geq L \mathcolor{blue}{\cdot} \min_k\{d\kappa_{q_k}\}$; \\
Define a new equally spaced grid for $q$: $q_0=q_{k^*}>q_1>\dots>q_K=q_{min}>0$; 
}
\Output{$q^*$}
\caption{Tuning the hyper-parameter $q$ with the function $\kappa$}
\label{algq2}
\end{algorithm}

In addition to the evaluation metric, another change we make here is that instead of the fixed threshold in Algorithm \ref{algq1}, we apply a floating threshold based on the values of $\kappa_{q_k}$, with a threshold coefficient $L$. In our simulation studies in Section \ref{sec:simulation}, we set the value of the coefficient $L$ equal to~$4$. 

In general, the results given by Algorithms \ref{algq1} and \ref{algq2} are quite similar, while the results from Algorithm \ref{algq2} are more stable than those of Algorithm \ref{algq1}, as will be shown in Section \ref{sec:simulation} through a comparison. However, the computation using Algorithm \ref{algq2} is much faster than Algorithm \ref{algq1} due to its simplicity. Therefore, in Sections \ref{sec:simulation} and \ref{sec:realdata}, in which we conduct numerical experiments on both simulated and real world datasets, we tune the hyper-parameter $q$ using Algorithm \ref{algq2}.

\subsection{Computations}
\label{sec:compute}
The computation of the ML$q$E is implemented in the software \texttt{ExaGeoStat} (\citealp{exageostat}). To make computation easier and more compatible with the environment in \texttt{ExaGeoStat}, while evaluating the $L_q$-likelihood, we first evaluate the log-likelihood and perform an additional transformation on the log-likelihood if $q\neq1$. 

From (\ref{density}), the log-likelihood for a realisation $\mathbf{Z}$ of length $n$ from a zero-mean Gaussian random field with covariance matrix $\mathbf{\Sigma}_{\mathcal{M}}$ parametrised by $\boldsymbol{\theta}$ can be expressed as
\begin{equation*}
    l(\mathbf{Z}; \boldsymbol{\theta})=-\frac{n}{2}\log(2\pi)-\frac{1}{2}\mathbf{Z}^\top\mathbf{\Sigma}_{\mathcal{M}}^{-1}\mathbf{Z}-\frac{1}{2}\log|\mathbf{\Sigma}_{\mathcal{M}}|.
\end{equation*}
Furthermore, from (\ref{lq}), the expression of the $L_q$-likelihood in this case is
\begin{equation*}
    L_q(\mathbf{Z}; \boldsymbol{\theta})=
    \begin{cases}
    l(\mathbf{Z}; \boldsymbol{\theta}),\quad\text{if}\,\, q=1; \\
    \left\{\exp[l(\mathbf{Z}; \boldsymbol{\theta})\times(1-q)]-1\right\}/(1-q),\quad\text{otherwise}.
    \end{cases}
\end{equation*}

The detailed Algorithm \ref{alg2} for computing the $L_q$ likelihood can be found in Appendix~\ref{sec:alg}. In addition, as \cite{lq} pointed out, for a fixed $q$, the ML$q$E $\hat{\boldsymbol{\theta}}$ would converge to $\boldsymbol{\theta}_0/q$ in probability; therefore, a correction of the final estimation result could be considered. However, these authors also mentioned that the numerical results after the correction are not promising, a claim that was further confirmed by our own experiments. Therefore, in all the numerical studies in this work, we directly consider the estimation results without correction to be the ML$q$E results.


\section{Simulation Study}
\label{sec:simulation}
In this section, we conduct simulation studies on the proposed method, in which we use the software \texttt{ExaGeoStat} to both generate the data and estimate the parameters. The optimisation algorithm we use for parameter estimation is \texttt{BOBYQA} (\citealp{powell2009bobyqa}), which is a bound-constrained algorithm without using derivatives embedded in \texttt{ExaGeoStat}. 

In each of the experiments, we generate synthetic data from a zero-mean Gaussian random field with Matérn covariance matrix $\boldsymbol{\Sigma}_{\mathcal{M}}$ parametrised by $\boldsymbol{\theta}$, with $n$ locations and $m$ replicates, and the level of contamination $r$. If $r=0$, then we are generating clean data without outliers, which is used in Section \ref{sec:nooutlier}; if $r>0$, which means that the synthetic data are contaminated by outliers and is used in Section \ref{sec:withoutliers}, then we generate the clean data first and add noises to parts of the data afterward. The detailed Algorithm \ref{alg3} is shown in Appendix~\ref{sec:alg}. In the rest of this section, we show the estimation results of the variance, range and smoothness parameters in each setting. In addition, we plot the estimated value of the function $\kappa$ in (\ref{functionf}), to evaluate the fitness of the model with the estimated parameters. Throughout this section and the supplementary materials, in the figures in which we show our experimental results, the ``function $\kappa$'' refers to the function $\kappa$ in (\ref{functionf}). Moreover, for all the simulation experiments in this section, we tune the value of $q$ using Algorithm~\ref{algq2}. 

\subsection{Clean Data}
\label{sec:nooutlier}
First of all, we performed some experiments to compare the performance of ML$q$E with MLE when there are no outliers in the simulated data.

\begin{figure}[h!]
    \centering
    \includegraphics[width=\linewidth]{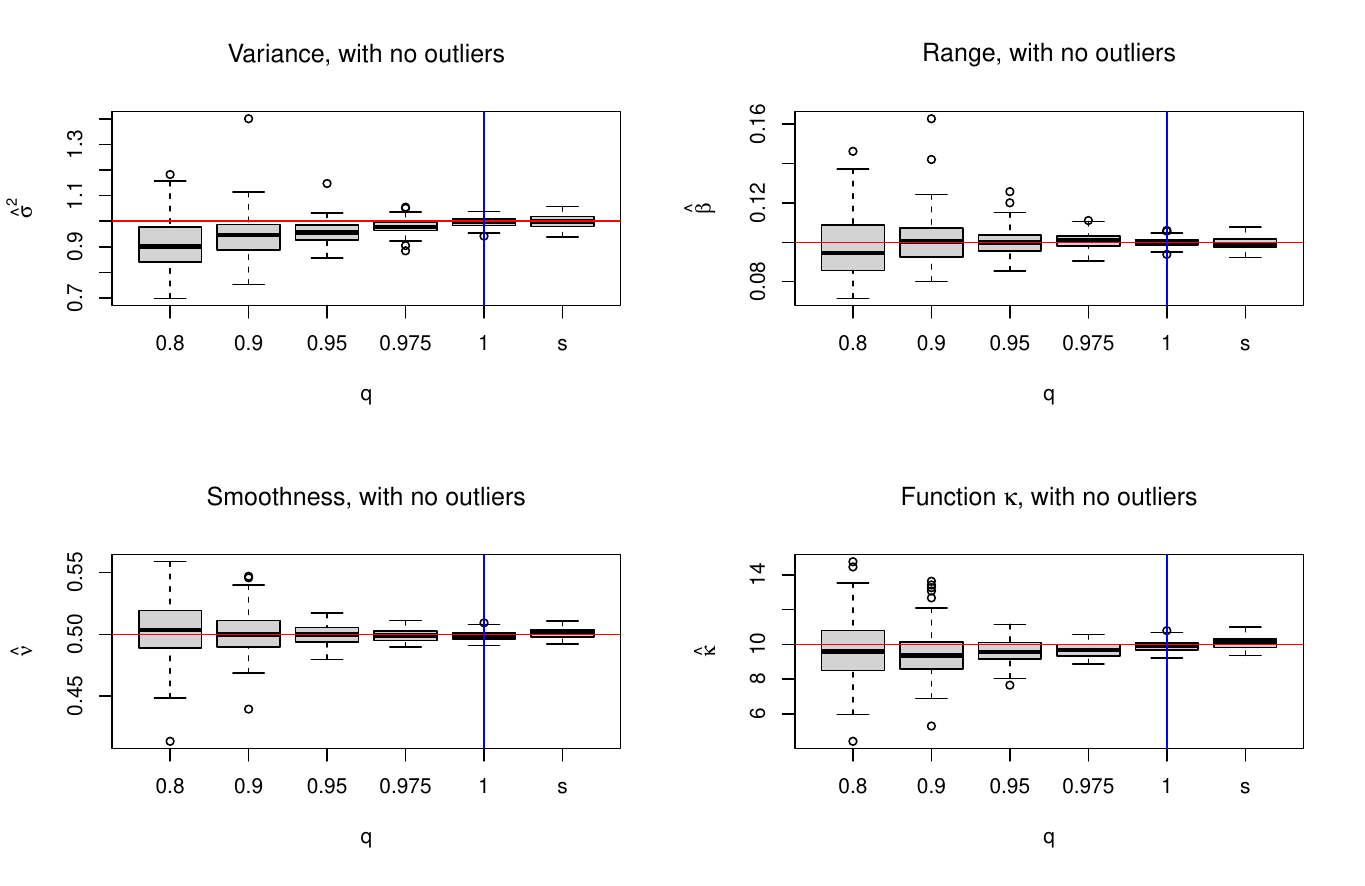}
    \caption{\footnotesize The MLE and ML$q$E estimation results with no outlier in the data. The red horizontal lines correspond to the true values of the parameters or the function $\kappa$ in (\ref{functionf}). The blue vertical lines correspond to the value of $q$ leading to the smallest mean squared error (MSE), which in this case is $q=1$ (MLE).
    }
    \label{nooutlier1}
\end{figure}

In Figure \ref{nooutlier1}, we show the experimental results with simulated data from the Matérn covariance function with $\sigma^2=1, \beta=0.1, \nu=0.5$, which is essentially the exponential covariance function with a medium spatial dependence strength. Here we simulate the data with $n=1,600$ locations and $m=100$ replicates, and the experiment is repeated on $100$ different datasets generated from the same random field to make the boxplots. We use red horizontal lines to indicate the true values of the parameters and the function $\kappa$ in (\ref{functionf}), and blue vertical lines correspond to the value of $q$ leading to the smallest mean squared error (MSE). We can see that in this case the MLE is giving the best performance in terms of both consistency and efficiency. We also notice that when $q$ is close to $1$, the ML$q$E is converging to the MLE as expected. 

However, in real applications, we do not have the luxury of performing parameter estimation on different datasets from the same random field to find the value of $q$ with the smallest MSE. Therefore, we also conduct choice-of-$q$ experiments using Algorithm \ref{algq2} for each of the $100$ datasets individually, to examine whether the optimal value of $q$ can also be found in this way. In the last column of each panel of Figure \ref{nooutlier1}, denoted by ``s'', we present the boxplot of the parameter estimation results using the values of $q$ chosen by Algorithm \ref{algq2}, which have slightly larger bias than the MLE results with $q=1$. This is also what we expect as the compromise of using a robust method, i.e., that it results in slightly larger variability than the MLE when applied on clean data.

\subsection{Contaminated Data}
\label{sec:withoutliers}
\begin{figure}[h!]
    \centering
    \includegraphics[width=1.1\linewidth]{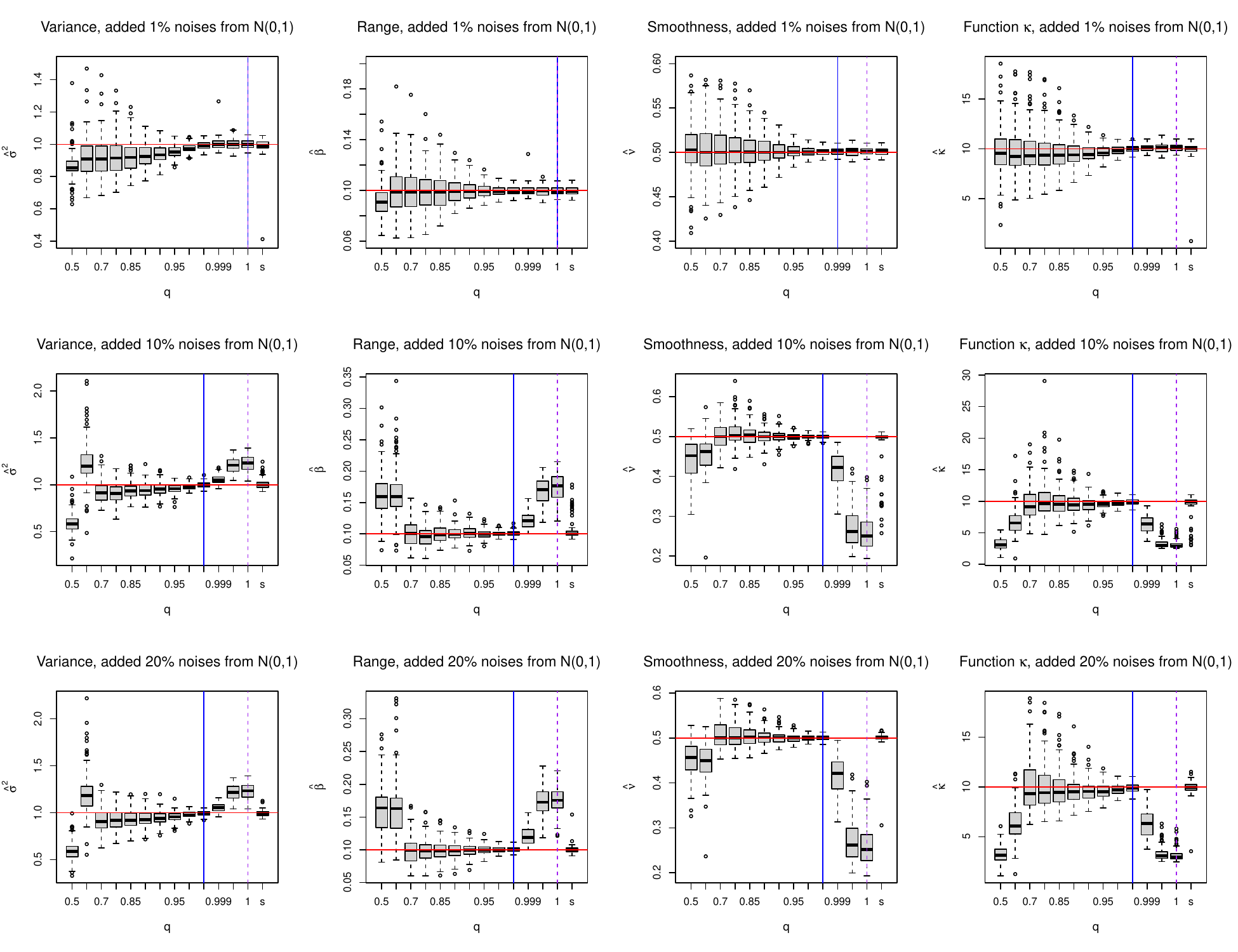}
    \caption{\footnotesize The MLE and ML$q$E estimation results with the data contaminated by noises generated from $N(0, 1)$, with $1\%$, $10\%$ and $20\%$ probability respectively from the first row to the third. The last column of each panel with the label ``s'' indicates the estimation results using the $q$ values selected using Algorithm \ref{algq2}. The values of $q$ we use here are $0.5$, $0.6$, $0.7$, $0.8$, $0.85$, $0.9$, $0.925$, $0.95$, $0.975$, $0.99$, $0.999$, $0.9999$, $1$. The red horizontal lines correspond to the true values of the parameters or the function $\kappa$ in (\ref{functionf}), and the blue vertical lines correspond to the value of $q$ leading to the smallest MSE. The MLE ($q=1$) results are indicated by the purple vertical dashed lines. In the first row, the $q$ values leading to the smallest MSE for variance and range are $1$, and in the corresponding figures the blue and purple lines overlap. 
    }
    \label{sd1}
\end{figure}

\begin{figure}[h!]
    \centering
    \includegraphics[width=1.1\linewidth]{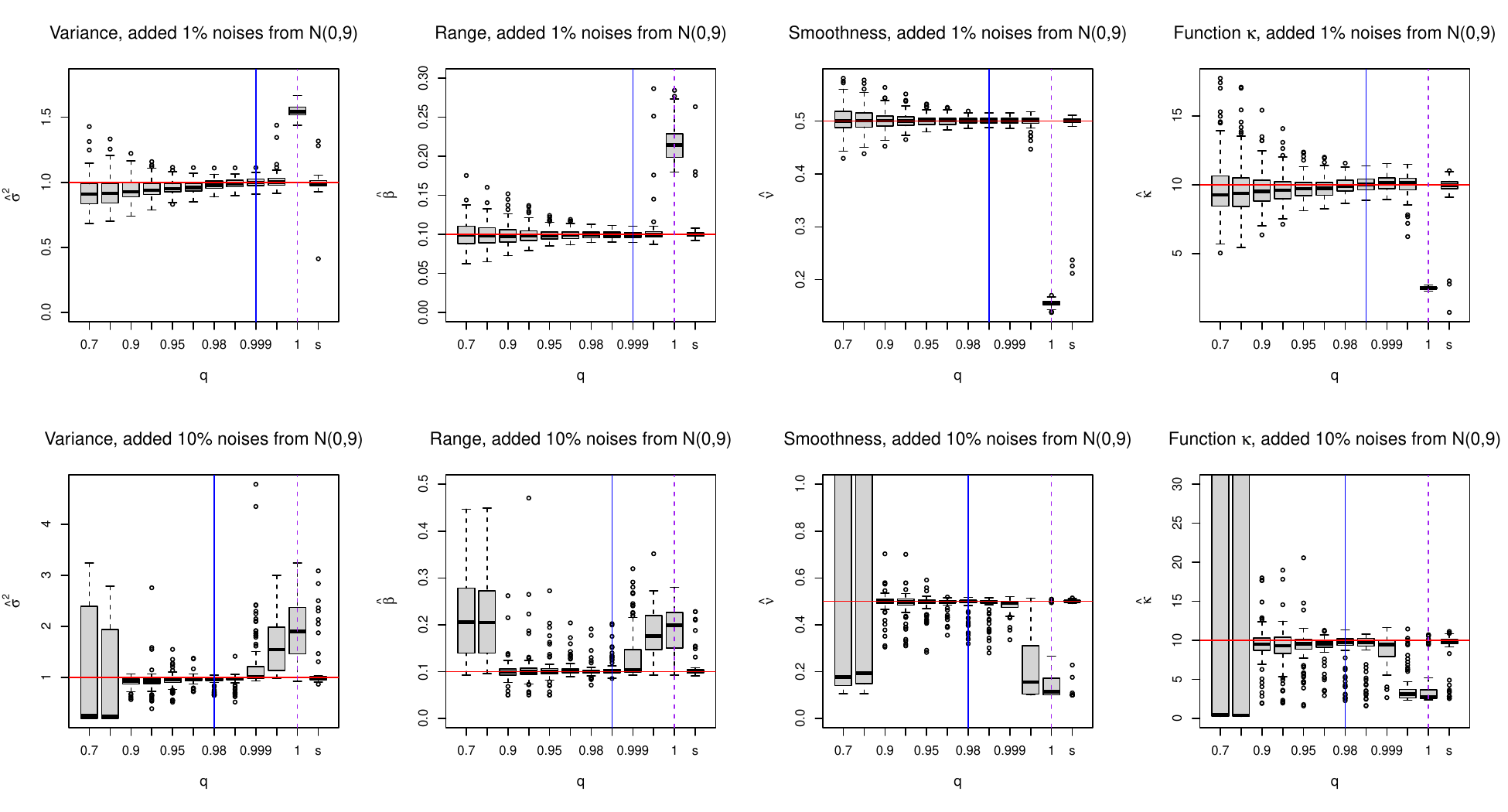}
    \caption{\footnotesize The MLE and ML$q$E estimation results with the data contaminated by noises generated from $N(0, 9)$, with $1\%$ probability in the first row and $10\%$ in the second. The last column of each panel with the label ``s'' indicates the estimation results using the $q$ values selected using Algorithm \ref{algq2}. The values of $q$ we use here are $0.7$, $0.8$, $0.9$, $0.925$, $0.95$, $0.96$, $0.98$, $0.99$, $0.999$, $0.9999$, $1$. The red horizontal lines correspond to the true values of the parameters or the function $\kappa$ in (\ref{functionf}), and the blue vertical lines correspond to the value of $q$ leading to the smallest MSE. The MLE ($q=1$) results are indicated by the purple vertical dashed lines. 
    }
    \label{sd3}
\end{figure}

Next, we test the performance of the ML$q$E and compare it with the MLE when there are different types of outliers in the simulated data. For all the results shown here in Figures \ref{sd1} and \ref{sd3}, the data are generated from a Gaussian random field with the Matérn covariance function with $\sigma^2=1, \beta=0.1, \nu=0.5$. Here also, we simulate the data with $n=1,600$ locations and $m=100$ replicates, and the experiment is repeated on $100$ different datasets generated from the same random field to make the boxplots. The mechanism of contaminating the data is as described in Algorithm \ref{alg3}. However, since we only have $100$ replicates for each dataset, in the experiments in which the data are contaminated with probability 1\% (see the first rows of Figures \ref{sd1} and \ref{sd3}) we actually randomly select exactly $1$ of the $100$ replicates to be contaminated.

In the first row of Figure \ref{sd1}, where we only add noises from $N(0, 1)$ to 1\% of the data, the MLE still performs well. For the variance and the range parameters, our hyper-parameter tuning method suggests that $q=1$ is the best choice. However, as we can see from the rest of Figures \ref{sd1} and \ref{sd3}, when we increase the proportion of data with noises from $N(0, 1)$ to 10\% or 20\%, the MLE results deviate significantly from the true values. Nevertheless, even if $q$ is only slightly smaller than $1$, the ML$q$E results are much closer to the true values, and $q=0.99$ is suggested to be the best choice for these two cases. For data with outliers generated by adding noises from $N(0, 9)$ with 1\% or 10\% probability, as we show in Figure \ref{sd3}, since the magnitude of outliers is much larger, smaller values of $q$ are needed to provide better ML$q$E results, compared with the data with added noises from $N(0, 1)$ with the same probability. In addition, as before, we use red horizontal lines to indicate the true values of the parameters and the function $\kappa$ in (\ref{functionf}), and blue vertical lines correspond to the value of $q$ leading to the smallest MSE. The MLE ($q=1$) results are indicated by the vertical purple dashed lines. 

Furthermore, similar to Figure \ref{nooutlier1}, in Figures \ref{sd1} and \ref{sd3}, we still add one column to the right of each panel labeled with ``s'', to present the boxplot of the parameter estimation results using the values of $q$ chosen by Algorithm \ref{algq2}. The choice-of-$q$ experiments are conducted on each of the $100$ datasets individually. We notice that in all three of these cases, for most simulated datasets, the estimation results using the chosen $q$ are similar to the results using the value of $q$ with the smallest values of MSE indicated by the blue vertical lines. In Figure \ref{sd3}, since outliers in the datasets are generated in a relatively extreme way, it is more often that the optimal value of $q$ is not chosen, compared to Figure \ref{sd1}. Nevertheless, despite those several extreme cases, Algorithm \ref{algq2} is working well and the resulting parameter estimation results are very close to the true values. 

Experimental results with synthetic datasets from some other values of the true parameters (i.e., other strengths of dependence and smoothness) are presented in Figures~S3 to S14 in the supplementary materials. In fact, the behaviour of the ML$q$E in those experiments is similar to the results shown here in Figures \ref{nooutlier1} to \ref{sd3}. For clean data, the MLE has the lowest bias and variability; the bias of the ML$q$E is also quite low as long as $q$ is not too small (generally not less than $0.95$), while it is not as good as the MLE in terms of efficiency. For contaminated data, the MLE is always biased for all cases, while the ML$q$E with $q$ between $0.97$ and $0.99$ generally gives fairly unbiased estimation results.  

\subsection{Chosen Values of $q$}
\begin{figure}[h!]
    \centering
    \includegraphics[width=\linewidth]{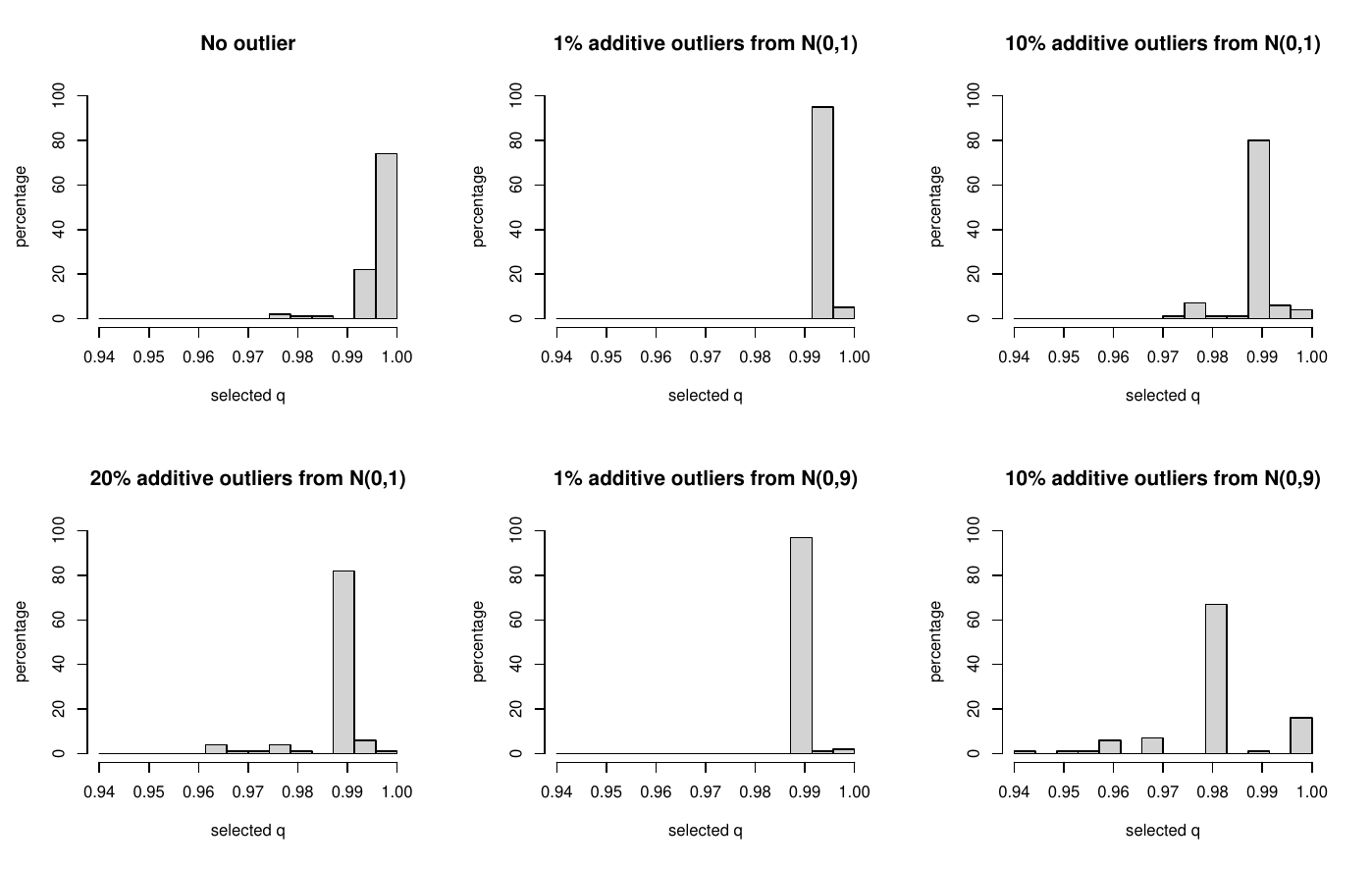}
    \caption{\footnotesize The histograms of the values of $q$ chosen by Algorithm \ref{algq2}, with the same synthetic data as in the experiments shown in Figures \ref{nooutlier1} to \ref{sd3}. The $y$-axes indicate the frequency (in percentage) of the selected $q$. 
    }
    \label{hist}
\end{figure}

In addition to the boxplots, we also present histograms of the frequency (in percentage) of the chosen values of $q$ using Algorithm \ref{algq2} in Figure \ref{hist}, with the same datasets that we used to plot Figures \ref{nooutlier1} to \ref{sd3}. 

For clean data, as expected, $q=1$ is chosen for most simulated datasets, and among the rest, $q>0.99$ is chosen for most cases, with $q<0.99$ chosen only a very small number of times. For contaminated data, in each of the five cases, the most frequently chosen value of $q$ is less than $1$, which means the ML$q$E outperforms the MLE. When the dataset is contaminated by noises from $N(0, 9)$, the most frequently chosen value of $q$ would be smaller than for the dataset contaminated by the same fraction of noises from $N(0, 1)$. Moreover, the algorithm to tune $q$ is designed in such a way that it chooses $1$ if the ML$q$E results are too unstable, which explains why in the most extreme case where the data contain 10\% outliers generated by adding noises from $N(0, 9)$, $q=1$ is chosen more frequently than the other cases of contaminated data. Those datasets for which $q=1$ is chosen also correspond to most of the outlying points in the columns labeled ``s'' in the second row of Figure \ref{sd3}.
\begin{figure}[t!]
    \centering
    \includegraphics[width=\linewidth]{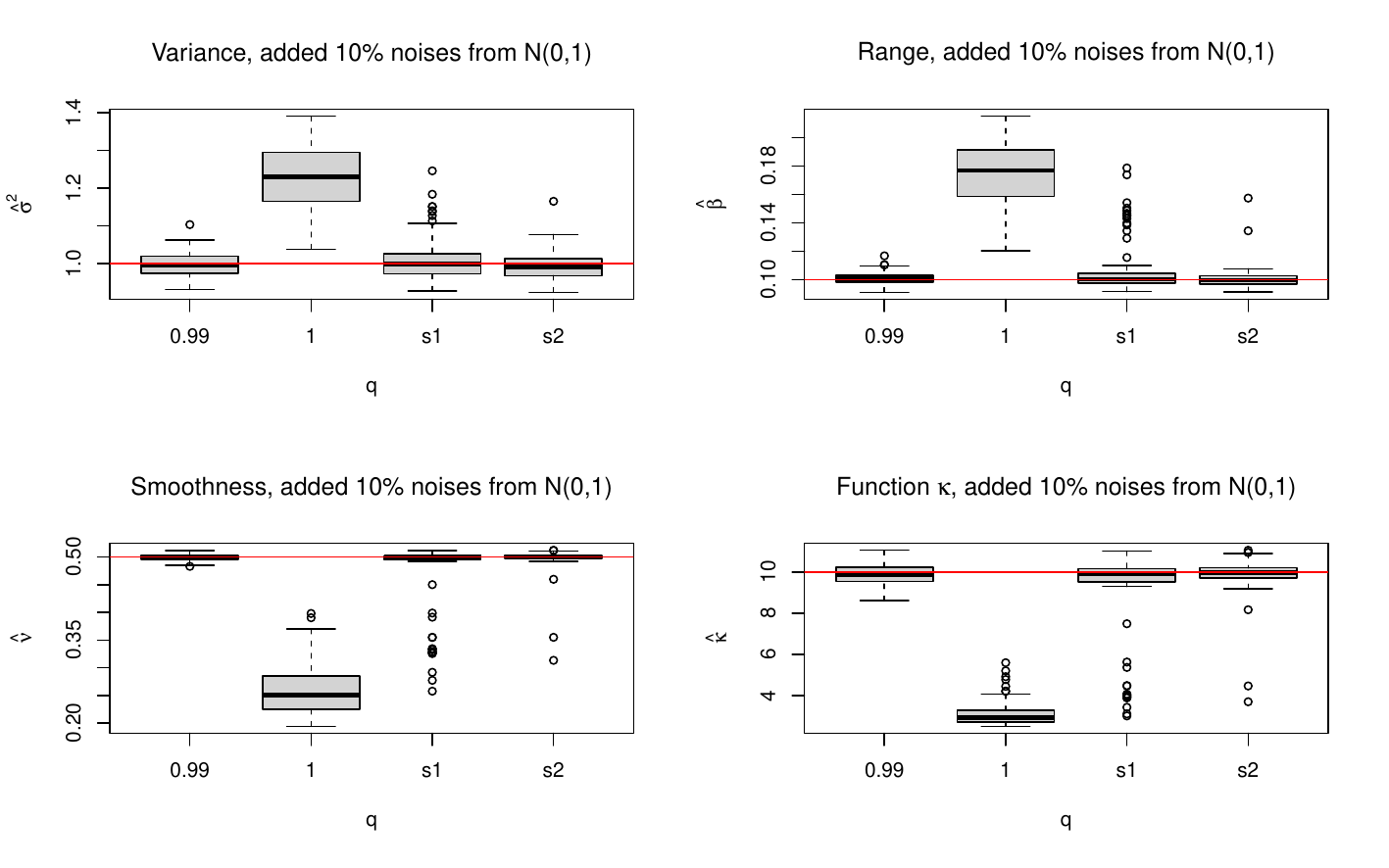}
    \caption{\footnotesize Comparison of the MLE and ML$q$E of several different cases. In each of the four sub-figures, from the left to the right, the four boxplots represent the ML$q$E of the smallest MSE in the second row of Figure \ref{sd1}, the MLE ($q=1$), the ML$q$E with $q$ selected by Algorithm \ref{algq1} (``s1''), and the ML$q$E with $q$ selected by Algorithm \ref{algq2} (``s2''), respectively. The red horizontal lines correspond to the true values of the parameters or the function $\kappa$ in (\ref{functionf}).
    }
    \label{compq}
\end{figure}

In addition, we present a comparison between the two methods of choosing $q$, namely Algorithms \ref{algq1} and \ref{algq2}, in Figure \ref{compq}, using the same data as the experiment shown in the second row of Figure \ref{sd1}. In Figure \ref{compq}, we show the parameter estimation results with the values of $q$ chosen by Algorithm \ref{algq1} (in the columns named ``s1'') and Algorithm \ref{algq2} (in the columns named ``s2''), together with the MLE results ($q=1$) and ML$q$E results with $q=0.99$, which is the $q$ leading to the smallest MSE in the second row of Figure \ref{sd1}. The red horizontal lines correspond to the true values of the parameters or the function $\kappa$ in (\ref{functionf}). Both Algorithms \ref{algq1} and \ref{algq2} are able to give quite unbiased estimation results, while the results in ``s2'' are more stable than in ``s1'', since there are many fewer outlying points. Further considering the massive advantage on computation speed of Algorithm \ref{algq2} over Algorithm~\ref{algq1}, in practice, it is  recommended to apply Algorithm \ref{algq2} to tune the hyper-parameter $q$.



\vspace{-.2cm}
\section{Precipitation Data Analysis}
\label{sec:realdata}
We return to the US precipitation data we used as a motivating example in Section \ref{sec:intro}. To eliminate the influence of missing data on our experiment, we select the data from 1928 to 1997 on $n=621$ observation stations whose data do not contain any missing values for those $m=70$ years. In addition, we extract the data of each month in the dataset and study them separately, making 12 sub-datasets. For example, for the experiment labeled ``Jan'', we used only the precipitation data of January from 1928 to 1997 at all observation stations. In this way, it is reasonable to assume that all the replicates (i.e., data from different years) in each experiment are from the same distribution. Moreover, to remove the trend in precipitation, the data for each year are centralised separately. 
\begin{figure}[t!]
    \centering
    \includegraphics[width=1.1\linewidth]{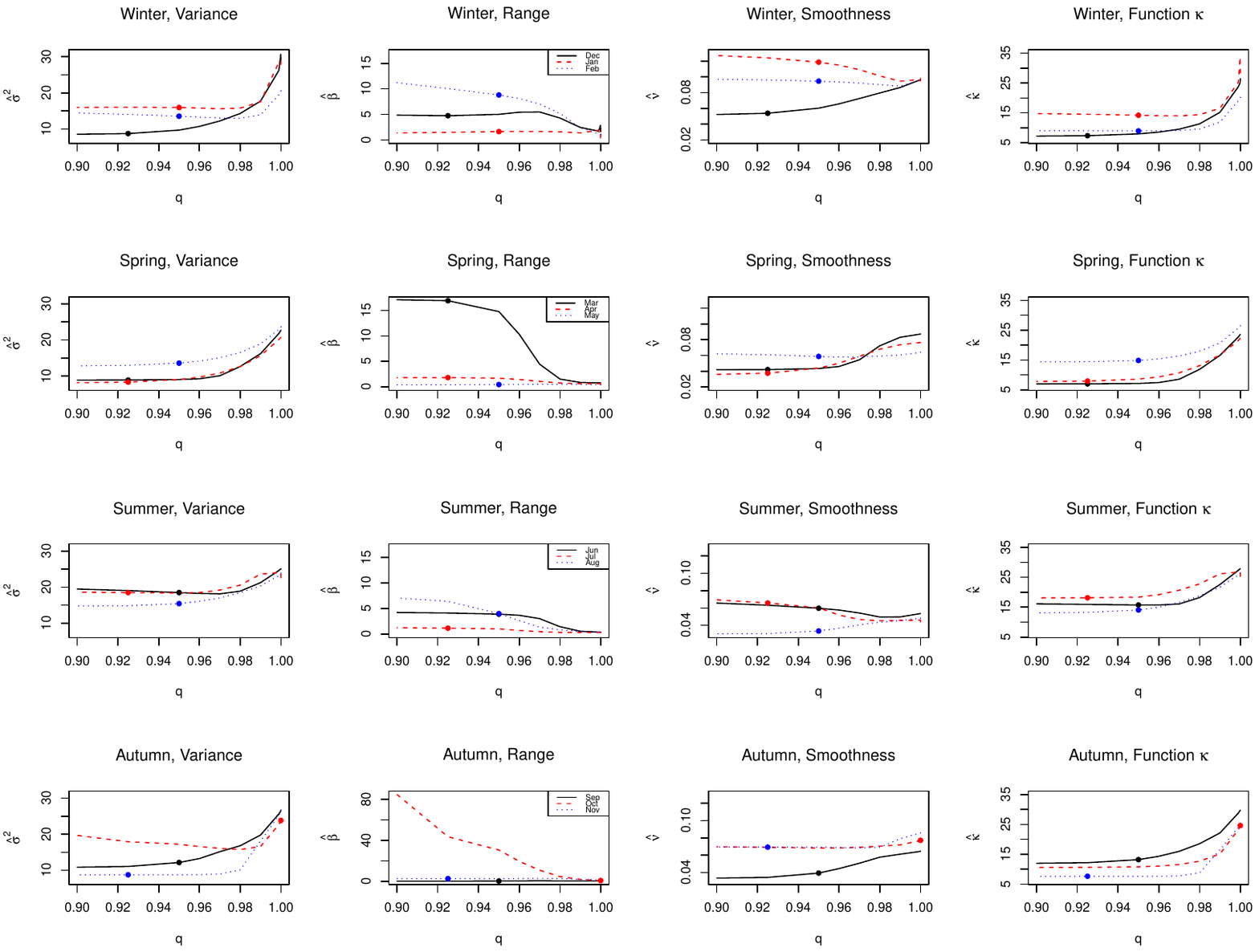}
    \caption{\footnotesize 
    The MLE and ML$q$E estimation results with different values of $q$, for the US precipitation data of different months. Each column shows the plot of the estimation of the same parameter or the function $\kappa$ in (\ref{functionf}), and each row shows the results for the same months. The legends in the second column of each row apply to all sub-figures in the same row. The dots on the curves indicate the values of $q$ selected with Algorithm \ref{algq2} and their corresponding estimation results. 
    }
    \label{real}
\end{figure}

In Figure \ref{real}, we present the experimental results using the data. In each of the subplots, the results corresponding to three consecutive months are presented for compactness. The triplets of the months are formed based on seasons: December, January and February correspond to Winter, March, April and May correspond to Spring, June, July and August correspond to Summer, and finally September, October and November correspond to Autumn. In the first 3 columns, we show the estimation results using ML$q$E with different values of $q$, while it degenerates to the MLE when $q=1$; in the last column, we show the estimated value of the function $\kappa$ in (\ref{functionf}). The values of $q$ that we use here are $0.9$, $0.925$, $0.95$, $0.96$, $0.97$, $0.98$, $0.99$, $0.999$, $0.9999$, and $1$. The dot on each curve shows the optimal value of $q$ selected for the data using Algorithm \ref{algq2}. 


We can see from Figure \ref{real} that, for all 12 months, the ML$q$E results with most values of $q<1$ are different from the corresponding MLE results. This observation indicates that outliers exist in the 12 sub-datasets of the 12 months; otherwise the ML$q$E results with $q$ slightly smaller than $1$ are expected to be similar to the MLE results. To examine whether there are really outliers in all the 12 sub-datasets, we apply the same method that we used to detect the outliers in the January data that we described in Section~\ref{sec:intro}, which is to first find the empirical variogram of the data from each year, and then find out the years for which the data behave differently from other years via functional boxplots using the \texttt{R} command \texttt{fbplot}. It turns out that outliers exist in all of the 12 sub-datasets, which explains why the MLE and ML$q$E results with $q<1$ are different for all the 12 months.
\begin{table}[h!]
\footnotesize 
\centering
\caption{The ML$q$E results of the three parameters of Matérn covariance function (\ref{matern}), $\sigma^2$, $\beta$ and $\nu$, for the US precipitation data of all the 12 months, using the values of $q$ selected via Algorithm \ref{algq2}, followed by the MLE results of the same data. \\}

\begin{tabular}{c|c|cc|cc|cc}
    & $q$ selected & $\hat{\sigma}^2$ & MLE     & $\hat{\beta}$ & MLE     & $\hat{\nu}$ & MLE     \\ \hline
Jan & $0.95$       & $15.94$          & $28.85$ & $1.614$       & $1.080$ & $0.119$     & $0.098$ \\
Feb & $0.95$       & $13.54$          & $20.57$ & $8.797$       & $1.099$ & $0.094$     & $0.096$ \\
Mar & $0.925$      & $8.908$          & $22.69$ & $16.92$       & $0.799$ & $0.042$     & $0.087$ \\
Apr & $0.925$      & $8.329$          & $20.73$ & $1.835$       & $0.621$ & $0.037$     & $0.076$ \\
May & $0.95$       & $13.59$          & $23.62$ & $0.467$       & $0.408$ & $0.059$     & $0.064$ \\
Jun & $0.95$       & $18.49$          & $25.09$ & $3.906$       & $0.371$ & $0.060$     & $0.053$ \\
Jul & $0.925$      & $18.61$          & $22.62$ & $1.230$       & $0.283$ & $0.069$     & $0.044$ \\
Aug & $0.95$       & $15.42$          & $23.92$ & $3.998$       & $0.360$ & $0.034$     & $0.048$ \\
Sep & $0.95$       & $12.16$          & $26.46$ & $0.354$       & $0.401$ & $0.039$     & $0.064$ \\
Oct & $1$          & $23.82$          & $23.82$ & $0.810$       & $0.810$ & $0.077$     & $0.077$ \\
Nov & $0.925$      & $8.765$          & $26.70$ & $2.712$       & $1.395$ & $0.069$     & $0.085$ \\
Dec & $0.925$      & $8.725$          & $28.93$ & $4.730$       & $1.580$ & $0.054$     & $0.095$
\end{tabular}
\label{rr}
\end{table}
In addition, we note that for the October data, the optimal selected value of $q$ is $1$, because the ML$q$E results, especially for the range parameter, cannot stabilise as $q$ decreases, and the MLE is considered as the optimal choice to ensure that we still have a reasonable estimation result. For the data for each of the other months, the optimal values of $q$ are all selected as a point such that the curve of the estimation results becomes relatively flat in a neighborhood of it, which means that stable ML$q$E results can be achieved using the values $q$ close to this value. We also present the ML$q$E results using the selected value of $q$, together with the MLE results, in Table \ref{rr}. Except for October, where the ML$q$E and MLE are identical because $q=1$ is chosen, the MLE overestimates the variance for all other months due to the presence of outliers in the datasets. Meanwhile, the MLE often underestimates both the range and smoothness of the Matérn covariance function. For the 11 months where the ML$q$E differs from the MLE, the range parameter is underestimated for 10 months and the smoothness is underestimated for 8 months. Outliers generally have higher variance compared to non-outliers, and the noise in the data reduces the range and smoothness of the random field. Consequently, the non-robust MLE tends to overestimate variance while underestimating range and smoothness. In contrast, the robustness of the ML$q$E provides estimates that more accurately reflect the true spatial dependence of the datasets. 


\section{Conclusion}
\label{sec:conclusion}
To conclude, both simulation and real data experiments have shown that the ML$q$E can indeed outperform the MLE with the existence of outliers in the data. With a suitable choice of $q$, the parameter estimation results via the ML$q$E exhibit very low bias and high efficiency, compared to the MLE, when the data are contaminated. On the theory side, the ML$q$E holds desirable asymptotic properties as well. The mechanism of tuning the hyper-parameter $q$ for the ML$q$E that we developed was also proved to work very well in our experiments. Moreover, our method can efficiently deal with data on a much larger scale than in the literature, with the help of the high-performance computing framework \texttt{ExaGeoStat}, which is extremely important for spatial data analysis.

A limitation of the current work is that the current method is only able to provide robust estimation for spatial data with multiple replicates, since it can only treat some of the replicates as outliers as a whole. An interesting follow-up work would be investigating how to adjust this method so that it can be applied on spatial data with only one replicate as well. Another possible extension of this work would be integrating the proposed method with high-performance approximation methods for numerical linear algebra, such as the tile low-rank algorithm and the mixed-precision method, so that the ML$q$E can be further applied on even larger-scale datasets efficiently. 

There are also several other possible extensions of the method presented in this work. First, for convenience, here we only considered the case in which all the replicates of the spatial data are from the same fixed set of locations. However, it is also possible to apply the proposed method on replicated spatial datasets where the locations for each of the replicates are not exactly the same, given that they are still from the same region so that we are able to assume that those replicated data are essentially from the same Gaussian random field. Second, we did not include the nugget effect in the model. It is also possible to estimate the nugget effect together with the three parameters in the Matérn covariance function that we considered, with some minor modifications of the code. However, for the choice-of-$q$ mechanism, since the nugget effect does not appear in the function $\kappa$ in (\ref{functionf}), Algorithm \ref{algq2} may not work that well in this case.
Furthermore, the methodology presented here can also be potentially adapted to estimate the parameters of non-Gaussian random fields, such as $t$-random fields \citep{roislien2006t} or Tukey $g$-and-$h$ random fields \citep{xu2017tukey}. These processes are useful to model non-Gaussian heavy-tailed or skewed spatial data. In the present work, however, we concentrated on the more popular Gaussian process with Matérn covariance function. In future work, the performance of this methodology for non-Gaussian processes can be explored.



\section*{Acknowledgements}
This publication is based upon work supported by King Abdullah University of Science and Technology Research Funding (KRF) under Award No. ORFS-2022-CRG11-5069.

\bibliographystyle{asa}
\bibliography{ref}

\appendix
\label{append}
\section{Proofs}
\label{sec:proof}

\subsection{Proof of Theorem \ref{th1}}
Note that
\begin{align*}
    \log f(\mathbf{z}; \boldsymbol{\theta})&=-\frac{n}{2}\log(2\pi)-\frac{1}{2}\log|\mathbf{\Sigma}_{\mathcal{M}}|-\frac{1}{2}\mathbf{z}^\top\mathbf{\Sigma}_{\mathcal{M}}^{-1}\mathbf{z}, \\
    \frac{\partial}{\partial\boldsymbol{\theta}}\log|\mathbf{\Sigma}_{\mathcal{M}}|&=\text{tr}\left(\mathbf{\Sigma}_{\mathcal{M}}^{-1}\frac{\partial}{\partial\boldsymbol{\theta}}\mathbf{\Sigma}_{\mathcal{M}}\right), \\
    \frac{\partial}{\partial\boldsymbol{\theta}}\mathbf{z}^\top\mathbf{\Sigma}_{\mathcal{M}}^{-1}\mathbf{z}&=\mathbf{z}^\top\left(\frac{\partial}{\partial\boldsymbol{\theta}}\mathbf{\Sigma}_{\mathcal{M}}^{-1}\right)\mathbf{z}=-\mathbf{z}^\top\mathbf{\Sigma}_{\mathcal{M}}^{-1}\frac{\partial\mathbf{\Sigma}_{\mathcal{M}}}{\partial\boldsymbol{\theta}}\mathbf{\Sigma}_{\mathcal{M}}^{-1}\mathbf{z} .
\end{align*}
Hence, we have
\begin{align*}
    \frac{\partial}{\partial\boldsymbol{\theta}}\log f(\mathbf{z}; \boldsymbol{\theta})
    &=-\frac{1}{2}\text{tr}\left(\mathbf{\Sigma}_{\mathcal{M}}^{-1}\frac{\partial}{\partial\boldsymbol{\theta}}\mathbf{\Sigma}_{\mathcal{M}}\right)+\frac{1}{2}\mathbf{z}^\top\mathbf{\Sigma}_{\mathcal{M}}^{-1}\frac{\partial\mathbf{\Sigma}_{\mathcal{M}}}{\partial\boldsymbol{\theta}}\mathbf{\Sigma}_{\mathcal{M}}^{-1}\mathbf{z}.
\end{align*}
Therefore, we let
\begin{align}
& \mathbf{U}^*_n( \mathbf{Z}; \boldsymbol{\theta}, q )=\frac{\partial}{\partial\boldsymbol{\theta}} L_q[f(\mathbf{Z}; \boldsymbol{\theta})] \nonumber= \frac{\partial}{\partial\boldsymbol{\theta}}\frac{f^{1-q}(\mathbf{Z}; \boldsymbol{\theta})-1}{1-q} \nonumber= f^{1-q}(\mathbf{Z}; \boldsymbol{\theta})\frac{\partial}{\partial\boldsymbol{\theta}}\log f(\mathbf{Z}; \boldsymbol{\theta}) \nonumber\\
=& \left(\frac{1}{(2\pi)^{\frac{n}{2}}|\mathbf{\Sigma}_{\mathcal{M}}|^{\frac{1}{2}}}\right)^{1-q}\exp\left(-\frac{1-q}{2}\left(\mathbf{Z}^\top\mathbf{\Sigma}_{\mathcal{M}}^{-1}\mathbf{Z}\right)\right)\left[\frac{1}{2}\mathbf{Z}^\top\mathbf{\Sigma}_{\mathcal{M}}^{-1}\frac{\partial\mathbf{\Sigma}_{\mathcal{M}}}{\partial\boldsymbol{\theta}}\mathbf{\Sigma}_{\mathcal{M}}^{-1}\mathbf{Z}-\frac{1}{2}\text{tr}\left(\mathbf{\Sigma}_{\mathcal{M}}^{-1}\frac{\partial}{\partial\boldsymbol{\theta}}\mathbf{\Sigma}_{\mathcal{M}}\right)\right] .
\label{thmeq1}
\end{align}
From \eqref{thmeq1} we get that the expression of $\mathbf{U}^*_n( \mathbf{Z}; \boldsymbol{\theta}, q )$, which is the same as (\ref{uv}) in the main text of this article, and it matches the quantity $\mathbf{U}^*( X; \theta, q )$ defined in (2.6) in \cite{lq}. Therefore, $\boldsymbol{\theta}_m^*$ defined here corresponds to $\theta_n^*$ defined in (3.2) in \cite{lq}.

Next,
\begin{align*}
    \frac{\partial^2}{\partial\boldsymbol{\theta}^2} L_q[f(\mathbf{Z}; \boldsymbol{\theta})]
    &= \left(\frac{\partial}{\partial\boldsymbol{\theta}}f^{1-q}(\mathbf{Z}; \boldsymbol{\theta})\right)\frac{\partial}{\partial\boldsymbol{\theta}}\log f(\mathbf{Z}; \boldsymbol{\theta}) + f^{1-q}(\mathbf{Z}; \boldsymbol{\theta})\frac{\partial^2}{\partial\boldsymbol{\theta}^2}\log f(\mathbf{Z}; \boldsymbol{\theta}) \\
    &= (1-q)f^{1-q}(\mathbf{Z}; \boldsymbol{\theta})\left(\frac{\partial}{\partial\boldsymbol{\theta}}\log f(\mathbf{Z}; \boldsymbol{\theta})\right)^2 + f^{1-q}(\mathbf{Z}; \boldsymbol{\theta})\frac{\partial^2}{\partial\boldsymbol{\theta}^2}\log f(\mathbf{Z}; \boldsymbol{\theta}).
\end{align*}
We proceed to derive the expression of $\frac{\partial^2}{\partial\boldsymbol{\theta}^2}\log f(\mathbf{Z}; \boldsymbol{\theta})$.
Note that
\begin{align*}
    \frac{\partial^2}{\partial\boldsymbol{\theta}^2}\log|\mathbf{\Sigma}_{\mathcal{M}}|&=\frac{\partial}{\partial\boldsymbol{\theta}}\text{tr}\left(\mathbf{\Sigma}_{\mathcal{M}}^{-1}\frac{\partial}{\partial\boldsymbol{\theta}}\mathbf{\Sigma}_{\mathcal{M}}\right)=\text{tr}\left(\frac{\partial}{\partial\boldsymbol{\theta}}\left(\mathbf{\Sigma}_{\mathcal{M}}^{-1}\frac{\partial}{\partial\boldsymbol{\theta}}\mathbf{\Sigma}_{\mathcal{M}}\right)\right) \\
    &=\text{tr}\left(\left(\frac{\partial}{\partial\boldsymbol{\theta}}\mathbf{\Sigma}_{\mathcal{M}}^{-1}\right)\left(\frac{\partial}{\partial\boldsymbol{\theta}}\mathbf{\Sigma}_{\mathcal{M}}\right)+\mathbf{\Sigma}_{\mathcal{M}}^{-1}\frac{\partial^2}{\partial\boldsymbol{\theta}^2}\mathbf{\Sigma}_{\mathcal{M}}\right), \\
    \frac{\partial}{\partial\boldsymbol{\theta}}\mathbf{z}^\top\mathbf{\Sigma}_{\mathcal{M}}\mathbf{z}&=\mathbf{z}^\top\left(\frac{\partial^2}{\partial\boldsymbol{\theta}^2}\mathbf{\Sigma}_{\mathcal{M}}^{-1}\right)\mathbf{z}=-\mathbf{z}^\top\left(\frac{\partial}{\partial\boldsymbol{\theta}}\left(\mathbf{\Sigma}_{\mathcal{M}}^{-1}\frac{\partial\mathbf{\Sigma}_{\mathcal{M}}}{\partial\boldsymbol{\theta}}\mathbf{\Sigma}_{\mathcal{M}}^{-1}\right)\right)\mathbf{z},\\
    \frac{\partial}{\partial\boldsymbol{\theta}}\left(\mathbf{\Sigma}_{\mathcal{M}}^{-1}\frac{\partial\mathbf{\Sigma}_{\mathcal{M}}}{\partial\boldsymbol{\theta}}\mathbf{\Sigma}_{\mathcal{M}}^{-1}\right)
    &=\left(\frac{\partial}{\partial\boldsymbol{\theta}}\mathbf{\Sigma}_{\mathcal{M}}^{-1}\right)\frac{\partial\mathbf{\Sigma}_{\mathcal{M}}}{\partial\boldsymbol{\theta}}\mathbf{\Sigma}_{\mathcal{M}}^{-1}+\mathbf{\Sigma}_{\mathcal{M}}^{-1}\frac{\partial}{\partial\boldsymbol{\theta}}\left(\frac{\partial\mathbf{\Sigma}_{\mathcal{M}}}{\partial\boldsymbol{\theta}}\mathbf{\Sigma}_{\mathcal{M}}^{-1}\right) \\
    &=\left(\frac{\partial}{\partial\boldsymbol{\theta}}\mathbf{\Sigma}_{\mathcal{M}}^{-1}\right)\frac{\partial\mathbf{\Sigma}_{\mathcal{M}}}{\partial\boldsymbol{\theta}}\mathbf{\Sigma}_{\mathcal{M}}^{-1}+\mathbf{\Sigma}_{\mathcal{M}}^{-1}\frac{\partial^2\mathbf{\Sigma}_{\mathcal{M}}}{\partial\boldsymbol{\theta}^2}\mathbf{\Sigma}_{\mathcal{M}}^{-1}+\mathbf{\Sigma}_{\mathcal{M}}^{-1}\frac{\partial\mathbf{\Sigma}_{\mathcal{M}}}{\partial\boldsymbol{\theta}}\left(\frac{\partial}{\partial\boldsymbol{\theta}}\mathbf{\Sigma}_{\mathcal{M}}^{-1}\right) \\
    &=-2\mathbf{\Sigma}_{\mathcal{M}}^{-1}\frac{\partial\mathbf{\Sigma}_{\mathcal{M}}}{\partial\boldsymbol{\theta}}\mathbf{\Sigma}_{\mathcal{M}}^{-1}\frac{\partial\mathbf{\Sigma}_{\mathcal{M}}}{\partial\boldsymbol{\theta}}\mathbf{\Sigma}_{\mathcal{M}}^{-1}+\mathbf{\Sigma}_{\mathcal{M}}^{-1}\frac{\partial^2\mathbf{\Sigma}_{\mathcal{M}}}{\partial\boldsymbol{\theta}^2}\mathbf{\Sigma}_{\mathcal{M}}^{-1} \\
    &=\mathbf{\Sigma}_{\mathcal{M}}^{-1}\left[\frac{\partial^2\mathbf{\Sigma}_{\mathcal{M}}}{\partial\boldsymbol{\theta}^2}-2\frac{\partial\mathbf{\Sigma}_{\mathcal{M}}}{\partial\boldsymbol{\theta}}\mathbf{\Sigma}_{\mathcal{M}}^{-1}\frac{\partial\mathbf{\Sigma}_{\mathcal{M}}}{\partial\boldsymbol{\theta}}\right]\mathbf{\Sigma}_{\mathcal{M}}^{-1},\\
    \frac{\partial^2}{\partial\boldsymbol{\theta}^2}\mathbf{z}^\top\mathbf{\Sigma}_{\mathcal{M}}\mathbf{z}
    &=-\mathbf{z}^\top\left(\frac{\partial}{\partial\boldsymbol{\theta}}\left(\mathbf{\Sigma}_{\mathcal{M}}^{-1}\frac{\partial\mathbf{\Sigma}_{\mathcal{M}}}{\partial\boldsymbol{\theta}}\mathbf{\Sigma}_{\mathcal{M}}^{-1}\right)\right)\mathbf{z} \\
    &=\mathbf{z}^\top\mathbf{\Sigma}_{\mathcal{M}}^{-1}\left[-\frac{\partial^2\mathbf{\Sigma}_{\mathcal{M}}}{\partial\boldsymbol{\theta}^2}+2\frac{\partial\mathbf{\Sigma}_{\mathcal{M}}}{\partial\boldsymbol{\theta}}\mathbf{\Sigma}_{\mathcal{M}}^{-1}\frac{\partial\mathbf{\Sigma}_{\mathcal{M}}}{\partial\boldsymbol{\theta}}\right]\mathbf{\Sigma}_{\mathcal{M}}^{-1}\mathbf{z}.
\end{align*}
Hence,
\begin{align*}
    \frac{\partial^2}{\partial\boldsymbol{\theta}^2}\log f(\mathbf{z}; \boldsymbol{\theta})
    =&-\frac{1}{2}\frac{\partial^2}{\partial\boldsymbol{\theta}^2}\log|\mathbf{\Sigma}_{\mathcal{M}}|-\frac{1}{2}\frac{\partial}{\partial\boldsymbol{\theta}}\mathbf{z}^\top\mathbf{\Sigma}_{\mathcal{M}}\mathbf{z} \\
    =&-\frac{1}{2}\text{tr}\left(\left(\frac{\partial}{\partial\boldsymbol{\theta}}\mathbf{\Sigma}_{\mathcal{M}}^{-1}\right)\left(\frac{\partial}{\partial\boldsymbol{\theta}}\mathbf{\Sigma}_{\mathcal{M}}\right)+\mathbf{\Sigma}_{\mathcal{M}}^{-1}\frac{\partial^2}{\partial\boldsymbol{\theta}^2}\mathbf{\Sigma}_{\mathcal{M}}\right) \\
    &-\frac{1}{2}\mathbf{z}^\top\mathbf{\Sigma}_{\mathcal{M}}^{-1}\left[-\frac{\partial^2\mathbf{\Sigma}_{\mathcal{M}}}{\partial\boldsymbol{\theta}^2}+2\frac{\partial\mathbf{\Sigma}_{\mathcal{M}}}{\partial\boldsymbol{\theta}}\mathbf{\Sigma}_{\mathcal{M}}^{-1}\frac{\partial\mathbf{\Sigma}_{\mathcal{M}}}{\partial\boldsymbol{\theta}}\right]\mathbf{\Sigma}_{\mathcal{M}}^{-1}\mathbf{z},
\end{align*}
and it follows that
\begin{align*}
    &\frac{\partial^2}{\partial\boldsymbol{\theta}^2}L_q[f(\mathbf{Z}; \boldsymbol{\theta})]  \\
    =& (1-q)f^{1-q}(\mathbf{Z}; \boldsymbol{\theta})\left(\frac{\partial}{\partial\boldsymbol{\theta}}\log f(\mathbf{Z}; \boldsymbol{\theta})\right)^2 + f^{1-q}(\mathbf{Z}; \boldsymbol{\theta})\frac{\partial^2}{\partial\boldsymbol{\theta}^2}\log f(\mathbf{Z}; \boldsymbol{\theta}) \\
    =& (1-q)\left(\frac{1}{(2\pi)^{\frac{n}{2}}|\mathbf{\Sigma}_{\mathcal{M}}|^{\frac{1}{2}}}\right)^{1-q}\exp\left(-\frac{1-q}{2}\left(\mathbf{Z}^\top\mathbf{\Sigma}_{\mathcal{M}}^{-1}\mathbf{Z}\right)\right)\\ &\times 
    \left[\frac{1}{2}\mathbf{Z}^\top\mathbf{\Sigma}_{\mathcal{M}}^{-1}\frac{\partial\mathbf{\Sigma}_{\mathcal{M}}}{\partial\boldsymbol{\theta}}\mathbf{\Sigma}_{\mathcal{M}}^{-1}\mathbf{Z}-\frac{1}{2}\text{tr}\left(\mathbf{\Sigma}_{\mathcal{M}}^{-1}\frac{\partial}{\partial\boldsymbol{\theta}}\mathbf{\Sigma}_{\mathcal{M}}\right)\right]^2 \\
    &+ \left(\frac{1}{(2\pi)^{\frac{n}{2}}|\mathbf{\Sigma}_{\mathcal{M}}|^{\frac{1}{2}}}\right)^{1-q}\exp\left(-\frac{1-q}{2}\left(\mathbf{Z}^\top\mathbf{\Sigma}_{\mathcal{M}}^{-1}\mathbf{Z}\right)\right)\\ &\times 
    \left\{\left[\frac{1}{2}\mathbf{Z}^\top\mathbf{\Sigma}_{\mathcal{M}}^{-1}\left[\frac{\partial^2\mathbf{\Sigma}_{\mathcal{M}}}{\partial\boldsymbol{\theta}^2}-2\frac{\partial\mathbf{\Sigma}_{\mathcal{M}}}{\partial\boldsymbol{\theta}}\mathbf{\Sigma}_{\mathcal{M}}^{-1}\frac{\partial\mathbf{\Sigma}_{\mathcal{M}}}{\partial\boldsymbol{\theta}}\right]\mathbf{\Sigma}_{\mathcal{M}}^{-1}\mathbf{Z}\right]\right.\\ &\left.-\frac{1}{2}\text{tr}\left(-\mathbf{\Sigma}_{\mathcal{M}}^{-1}\frac{\partial\mathbf{\Sigma}_{\mathcal{M}}}{\partial\boldsymbol{\theta}}\mathbf{\Sigma}_{\mathcal{M}}^{-1}\frac{\partial\mathbf{\Sigma}_{\mathcal{M}}}{\partial\boldsymbol{\theta}}+\mathbf{\Sigma}_{\mathcal{M}}^{-1}\frac{\partial^2\mathbf{\Sigma}_{\mathcal{M}}}{\partial\boldsymbol{\theta}^2}\right)\right\}.
\end{align*}
Therefore, the expressions of ${\bf K}_m$ and ${\bf J}_m$ defined in the statement of the theorem match with $K_n$ and $J_n$ defined in (3.4) and (3.5) in \cite{lq}, respectively, which completes the proof with reference to Theorem~3.2 and Theorem 4.3 in \cite{lq}.

\subsection{Expressions of Derivatives}
For practical use of Theorem \ref{th1}, we need to evaluate $\frac{\partial}{\partial\boldsymbol{\theta}}\mathbf{\Sigma}_{\mathcal{M}}$ and $\frac{\partial^2}{\partial\boldsymbol{\theta}^2}\mathbf{\Sigma}_{\mathcal{M}}$. Note that $\frac{\partial}{\partial\boldsymbol{\theta}}\mathbf{\Sigma}_{\mathcal{M}}$ and $\frac{\partial^2}{\partial\boldsymbol{\theta}^2}\mathbf{\Sigma}_{\mathcal{M}}$ are arrays consisting of the second-order derivatives of the terms of $\mathbf{\Sigma}_{\mathcal{M}}$ with respect to $\boldsymbol{\theta}$. The terms of $\mathbf{\Sigma}_{\mathcal{M}}$ are of the form as in (\ref{matern}).

Then we have
\begin{align*}
    \frac{\partial \mathcal{M}(h; \boldsymbol{\theta})}{\partial\sigma^2}=&\frac{1}{\Gamma(\nu)2^{\nu-1}}\left(\frac{h}{\beta}\right)^\nu \mathcal{K}_\nu\left(\frac{h}{\beta}\right); \\
    \frac{\partial^2\mathcal{M}(h; \boldsymbol{\theta})}{\partial(\sigma^2)^2}=&0; \\
    \frac{\partial \mathcal{M}(h; \boldsymbol{\theta})}{\partial\beta}=&\frac{\sigma^2}{\Gamma(\nu)2^{\nu-1}}\left\{-\frac{\nu}{\beta}\left(\frac{h}{\beta}\right)^\nu \mathcal{K}_\nu\left(\frac{h}{\beta}\right)-\left(\frac{h}{\beta}\right)^\nu \mathcal{K}'_\nu\left(\frac{h}{\beta}\right)\frac{h}{\beta^2}\right\}; \\
    \frac{\partial^2\mathcal{M}(h; \boldsymbol{\theta})}{\partial\beta^2}=&\frac{\sigma^2}{\Gamma(\nu)2^{\nu-1}}\left\{\frac{h}{\beta^2}\nu\left(\frac{h}{\beta}\right)^{\nu-1}\left[\frac{\nu}{\beta}\mathcal{K}_\nu\left(\frac{h}{\beta}\right)+\mathcal{K}'_\nu\left(\frac{h}{\beta}\right)\frac{h}{\beta}\right]\right. \\ &\left.-\left(\frac{h}{\beta}\right)^\nu\left[-\frac{\nu}{\beta^2}\mathcal{K}_\nu\left(\frac{h}{\beta}\right)-\frac{\nu}{\beta}\frac{h}{\beta^2}\mathcal{K}'_\nu\left(\frac{h}{\beta}\right)-\frac{h}{\beta^2}\mathcal{K}'_\nu\left(\frac{h}{\beta}\right)-\frac{h^2}{\beta^3}\mathcal{K}''_\nu\left(\frac{h}{\beta}\right)\right]\right\}; \\
    \frac{\partial \mathcal{M}(h; \boldsymbol{\theta})}{\partial\nu}=&\frac{\sigma^2}{\Gamma(\nu)2^{\nu-1}}\left\{-\left(\log(2)+\Psi(\nu)\right)\left(\frac{h}{\beta}\right)^\nu \mathcal{K}_\nu\left(\frac{h}{\beta}\right)+\frac{\partial}{\partial\nu}\left[\left(\frac{h}{\beta}\right)^\nu \mathcal{K}_\nu\left(\frac{h}{\beta}\right)\right]\right\}; \\
    \frac{\partial^2\mathcal{M}(h; \boldsymbol{\theta})}{\partial\nu^2}=&-\left(\log(2)+\Psi(\nu)\right)\frac{\sigma^2}{\Gamma(\nu)2^{\nu-1}}\\ &\times\left\{-\left(\log(2)+\Psi(\nu)\right)\left(\frac{h}{\beta}\right)^\nu \mathcal{K}_\nu\left(\frac{h}{\beta}\right)+\frac{\partial}{\partial\nu}\left[\left(\frac{h}{\beta}\right)^\nu \mathcal{K}_\nu\left(\frac{h}{\beta}\right)\right]\right\} \\
    &-\Psi'(\nu)\frac{\sigma^2}{\Gamma(\nu)2^{\nu-1}}\left(\frac{h}{\beta}\right)^\nu \mathcal{K}_\nu\left(\frac{h}{\beta}\right)\\
    &+\frac{\sigma^2}{\Gamma(\nu)2^{\nu-1}}\left\{-\left(\log(2)+\Psi(\nu)\right)\frac{\partial}{\partial\nu}\left[\left(\frac{h}{\beta}\right)^\nu \mathcal{K}_\nu\left(\frac{h}{\beta}\right)\right]+\frac{\partial^2}{\partial\nu^2}\left[\left(\frac{h}{\beta}\right)^\nu \mathcal{K}_\nu\left(\frac{h}{\beta}\right)\right]\right\}.
\end{align*}
Here, $\Psi(\cdot)$ represents the digamma function.

For the cross terms:
\begin{align*}
    \frac{\partial^2\mathcal{M}(h; \boldsymbol{\theta})}{\partial(\sigma^2)\partial\beta}=&\frac{1}{\Gamma(\nu)2^{\nu-1}}\left\{-\frac{\nu}{\beta}\left(\frac{h}{\beta}\right)^\nu \mathcal{K}_\nu\left(\frac{h}{\beta}\right)-\left(\frac{h}{\beta}\right)^\nu \mathcal{K}'_\nu\left(\frac{h}{\beta}\right)\frac{h}{\beta^2}\right\}; \\
    \frac{\partial^2\mathcal{M}(h; \boldsymbol{\theta})}{\partial(\sigma^2)\partial\nu}=&\frac{1}{\Gamma(\nu)2^{\nu-1}}\left\{-\left(\log(2)+\Psi(\nu)\right)\left(\frac{h}{\beta}\right)^\nu \mathcal{K}_\nu\left(\frac{h}{\beta}\right)+\frac{\partial}{\partial\nu}\left[\left(\frac{h}{\beta}\right)^\nu \mathcal{K}_\nu\left(\frac{h}{\beta}\right)\right]\right\}; \\
    \frac{\partial^2\mathcal{M}(h; \boldsymbol{\theta})}{\partial\beta\partial\nu}=&\frac{\sigma^2}{\Gamma(\nu)2^{\nu-1}}\left\{\left(\log(2)+\Psi(\nu)\right)\left[\frac{\nu}{\beta}\left(\frac{h}{\beta}\right)^\nu \mathcal{K}_\nu\left(\frac{h}{\beta}\right)+\left(\frac{h}{\beta}\right)^\nu \mathcal{K}'_\nu\left(\frac{h}{\beta}\right)\frac{h}{\beta^2}\right]\right. \\
    &-\left.\frac{1}{\beta}\left(\frac{h}{\beta}\right)^\nu \mathcal{K}_\nu\left(\frac{h}{\beta}\right)-\frac{\nu}{\beta}\frac{\partial}{\partial\nu}\left[\left(\frac{h}{\beta}\right)^\nu \mathcal{K}_\nu\left(\frac{h}{\beta}\right)\right]-\frac{h}{\beta^2}\frac{\partial}{\partial\nu}\left[\left(\frac{h}{\beta}\right)^\nu \mathcal{K}'_\nu\left(\frac{h}{\beta}\right)\right]\right\}.
\end{align*}

\section{Algorithms}
\label{sec:alg}

Here in Algorithm \ref{alg2}, we describe how we evaluate the $L_q$-likelihood for an observation vector $\mathbf{Z}$ from a zero-mean Gaussian random field with Matérn covariance parametrised by $\boldsymbol{\theta}$. The Cholesky factorisation, solving of the linear system and dot product are all parallelised in \texttt{ExaGeoStat}.
\begin{algorithm}[h!]
\footnotesize
\SetAlgoLined
\SetKwInOut{Input}{Input}
\SetKwInOut{Output}{Output}
\smallskip
\Input{Observation vector $\mathbf{Z}$ of length $n$, parameter $\boldsymbol{\theta}$ (initial value or the value from the previous iteration of optimisation), the $n$ locations}
\smallskip
Calculate the Matérn covariance matrix $\boldsymbol{\Sigma}_{\mathcal{M}}$ using $\boldsymbol{\theta}$ and the locations based on (\ref{matern}); \\
$\mathbf{L}\mathbf{L}^\top=\boldsymbol{\Sigma}_{\mathcal{M}}$: Cholesky factorisation; \\
Calculate $\mathbf{Z}_{new}$ by solving the linear system $\mathbf{L}\mathbf{L}^\top\mathbf{Z}_{new}=\mathbf{Z}$; \\
$logdet \gets \log|\boldsymbol{\Sigma}_{\mathcal{M}}|$; \\
$prod \gets \mathbf{Z}_{new}^\top\mathbf{Z}_{new}$; \\
$lq \gets - 0.5\cdot prod - 0.5\cdot logdet - 0.5\cdot n\cdot\log(2\pi)$; \\
\If{$q\neq1$} {
\smallskip
$lq\gets\{\exp[lq\times(1-q)]-1\}/(1-q)$; \\
}
\Output{The $L_q$ likelihood value $lq$}
\caption{Calculation of the $L_q$-likelihood}
\label{alg2}
\end{algorithm}
\medskip

In practice, since $q$ is fixed throughout one single optimisation process, when $q\neq1$, to simplify the calculation in the final step of the algorithm, we only need to let $lq\gets\exp[lq\times(1-q)]$ instead of evaluating the full expression. Moreover, since the value of the log-likelihood can be very small, the term $\exp[lq\times(1-q)]$ is likely to be exactly $0$ in most computing systems when the number of locations $n$ gets large. To solve this issue, again in the final step of the algorithm, when $q\neq1$, we let $lq\gets\exp[(lq+n)\times(1-q)]$, which means that we multiply the $L_q$-likelihood by the constant $\exp[n\times(1-q)]$, to make the computation possible. 

In Algorithm \ref{alg3}, we describe how we generate synthetic data from a zero-mean Gaussian random field with Matérn covariance matrix $\boldsymbol{\Sigma}_{\mathcal{M}}$ parametrised by $\boldsymbol{\theta}$, with $n$ locations and $m$ replicates, and level of contamination $r$. 
\begin{algorithm}[h!]
\footnotesize
\SetAlgoLined
\SetKwInOut{Input}{Input}
\SetKwInOut{Output}{Output}
\smallskip
\Input{Fixed locations, true parameter $\boldsymbol{\theta}$, number of locations $n$, number of replicates $m$, level of contamination $r$ with $0\leq r<1$}
\smallskip
Calculate the Matérn covariance matrix $\boldsymbol{\Sigma}_{\mathcal{M}}$ using $\boldsymbol{\theta}$ and the locations, based on (\ref{matern}); \\
$\mathbf{L}\mathbf{L}^\top=\boldsymbol{\Sigma}_{\mathcal{M}}$: Cholesky factorisation; \\
\For{i in $1:m$} {
\smallskip
    Generate vector $\mathbf{e}_i$ of length $n$ from i.i.d. standard normal distribution; \\
    $\mathbf{Z}_i\gets\mathbf{L}\cdot\mathbf{e}_i$; \\
    \If{$r>0$} {
    \smallskip
        Generate $r_i$ from continuous uniform distribution between $0$ and $1$; \\
        \If{$r_i<r$} {
        \smallskip
            Generate random noise vector $\mathbf{z}_i$ of length $n$ from i.i.d. normal (or other reasonable distributions); \\
            $\mathbf{Z}_i\gets\mathbf{Z}_i+\mathbf{z}_i$; \\
        }
    }
}
\Output{realisations $\{\mathbf{Z}_i, i=1, \dots, m\}$}
\caption{Generating synthetic data for simulation experiments}
\label{alg3}
\end{algorithm}

\end{document}